\documentclass[12pt]{iopart}
\usepackage{graphicx}
\usepackage{url}
\usepackage{mathrsfs}
\newcommand{\BE}{\begin{equation}}
\newcommand{\EE}{\end{equation}}
\expandafter\let\csname equation*\endcsname\relax
\expandafter\let\csname endequation*\endcsname\relax
\usepackage{amsmath,amssymb,amscd,latexsym,amsthm,mathrsfs}
\def\key{(\ref{eq:key}) }
\def\V{{\mathcal V}}
\def\N{{\mathcal N}}
\def\e{{\rm e}}
\def\m{{\rm m}}

\usepackage{hyperref}
\usepackage{subcaption}
\usepackage{ tikz}
\begin{document}

\title {\Large Statistical Mechanics of Confined Polymer Networks}
%\vskip 0.3cm

\author{Bertrand Duplantier$^\dag$, Anthony J Guttmann$^\ddag$}
\address{$^\dag$ Universit\'e Paris-Saclay, CNRS, CEA, Institut de Physique Th\'eorique, 91191, Gif-sur-Yvette, {\sc France}}
\address{$^\ddag$ School of Mathematics and Statistics,
The University of Melbourne, Victoria 3010, {\sc Australia}}

\setcounter{footnote}{0}

\begin{abstract}
We show how the theory of the critical behaviour of $d$-dimensional polymer networks of arbitrary topology can be generalized to the case of networks \emph{confined by hyperplanes}. This in particular encompasses  the case of a single polymer chain in a \emph{bridge} configuration. We further define \emph{multi-bridge} networks, where several vertices are in local bridge configurations.  We consider all cases of ordinary, mixed and special surface transitions, and polymer chains made of self-avoiding walks, or of mutually-avoiding walks, or at the tricritical $\Theta$-point. In the $\Theta$-point case, generalising the good-solvent case, we relate the critical exponent for simple bridges, $\gamma_b^{\Theta}$, to that of terminally-attached arches, $\gamma_{11}^{\Theta},$ and to the correlation length exponent $\nu^{\Theta}.$ We find $\gamma_b^{\Theta} = \gamma_{11}^{\Theta}+\nu^{\Theta}.$ In the case of the special transition, we find 
$\gamma_b^{\Theta}({\rm sp}) = \frac{1}{2}[\gamma_{11}^{\Theta}({\rm sp})+\gamma_{11}^{\Theta}]+\nu^{\Theta}.$ For general networks, the explicit expression of configurational exponents then naturally involve bulk and surface exponents for multiple random paths. In two-dimensions, we  describe their Euclidean exponents from a unified perspective, using Schramm-Loewner Evolution (SLE) in Liouville quantum gravity (LQG), and the so-called KPZ relation between Euclidean and LQG scaling dimensions. This is done  in the case of ordinary, mixed and special surface transitions, and of the $\Theta$-point. We provide compelling numerical evidence for some of these results both in two- and three-dimensions. %\textcolor{blue}{Another subset of these $\Theta$-walks, called {\em worms}, are defined as the subset of walks whose origin and end-point have the same $x$-coordinate. We give a scaling relation for the corresponding critical exponent $\gamma_w^{\Theta},$ which is $\gamma_w^{\Theta}=\gamma^{\Theta}-\nu^{\Theta}.$ This too is supported by enumerative results in the two-dimensional case. Using a direct renormalisation theory for polymers, we also study the cases of the upper-critical dimensions 4 for SAWs and 3 for the $\Theta$-point, including all logarithmic corrections to the corresponding Brownian polymer network partition functions.} 

\end{abstract}

 \vspace{5mm}

In celebration of the achievements of our dear colleague {\sc Joel L. Lebowitz.}\\

\section{Introduction}
\subsection{A brief history} 
Long polymer chains in a good solvent can be modelled in the continuum by the celebrated Edwards model \cite{Edwards65}, or, in a discrete setting, as self-avoiding walks (SAWs) on a lattice. It is well-known that they constitute
a \emph{critical} system.   This was originally recognized in a breakthrough paper by  P.-G. de Gennes \cite{PGG72} (which was part of his 1991 Nobel Prize in Physics) through their equivalence
to a magnetic $n$-component spin model, with $O(n)$ symmetry, in the
limit $n \to 0$. This allowed him to obtain the \emph{size and configuration critical exponents $\nu$ and $\gamma$}  of a single polymer chain, such that for $N\to +\infty$,
\begin{align}\label{eq:nu}
&R^2\propto N^{2\nu},
\\ \label{eq:gamma}
&\mathcal Z \propto {\mu}^N N^{\gamma-1},
\end{align}
where $R^2$ is the averaged square end-to-end distance of a  chain of $N$ 
monomers, and $\mathcal Z$ its partition function in a continuum model, or its self-avoiding configuration number in the lattice setting (up to translations) \cite{Flory,PGG79}. The constant $\mu$ is the \emph{connective constant}, or growth constant which is model-dependent, and whose   existence  was established by Hammersley  in the lattice case \cite{Hammer57}.  
In the magnetic $n$-component model, $\nu$ and $\mu$ are the universal correlation length
and susceptibility exponents, which depend solely on $n$ and the space dimension $d$. Scaling expressions such as \eqref{eq:nu} and \eqref{eq:gamma} are expected to hold up to fixed but non-universal amplitude factors. The  $\varepsilon$-expansions of $\nu,\gamma$ in space
dimension $d= 4- \varepsilon$ were then obtained in the famous Wilson-Fisher renormalization group  approach \cite{WF72} to the $(\varphi^2)^2$ interacting theory of a $n$-component field $\varphi$, taken in the $n \to 0$ limit in the polymer case \cite{PGG72}. This original field theoretic approach  was successfully extended to the case of \emph{polymer solutions} by J. des Cloizeaux \cite{JdC75} (see also \cite{SW77,SW80,WS78}), who later invented a \emph{direct renormalization method} \cite{JdC80,JdC81,JdC89} for the canonical Edwards model, which had the great advantage of being both geometrically intuitive and efficient, and was shown to be equivalent to that of field theory \cite{BM86,BD86b}.  The so-called Flory $\Theta$-point, where the solvent quality suddenly decreases  \cite{Flory}, corresponds to a \emph{tricritical} phase transition \cite{PGG75} and to a $(\varphi^2)^3$ interacting theory of a $n$-component field $\varphi$, still in the $n \to 0$ limit \cite{PGG78,BD82}. (For a recent mathematical study, see \cite{BauSla2019,BLS20}). The corresponding canonical Edwards model with 3-body interactions is itself amenable to a direct renormalization method \cite{BD86a,BM86,BD86c,JdC89}. Extensive Neutron and Light scattering experiments were performed on polymer solutions in the 70's, which neatly confirmed  scaling and renormalization theories \cite{Daoudetal75,DJ76}. Comprehensive monographs are \cite{JdC89,LS99}. 

In the case of two-dimensional self-avoiding walks, a breakthrough occurred in 1982, when B. Nienhuis \cite{Nienhuis82,Nienhuis87} (see also \cite{CardyHamber80}) used a mapping of the $O(n)$ model on the hexagonal lattice to a Coulomb gas to predict the exact (albeit non-rigorous) values of the critical exponents,
\begin{equation}\label{nugamma}
\nu=3/4,\quad \gamma=43/32.
\end{equation}
Another unexpected approach came in 1988 from so-called two-dimensional quantum gravity when Knizhnik, Polyakov and Zamolodchikov discovered the celebrated KPZ relation between critical exponents in the plane and those on a random lattice \cite{KPZ88,KPZbis,DKaw89}, via the use of \emph{Liouville quantum gravity} (LQG) \cite{Polya81}. It allowed in particular another derivation of SAW's exponents \eqref{nugamma} from a direct computation on a random lattice \cite{DK88,DK90}. 

Finally, one should mention  the groundbreaking invention in 1999  of \emph{Stochastic Loewner Evolution} (SLE)  by Oded Schramm \cite{Schramm2000}, which was a game-changer for the mathematical approach to the two-dimensional statistical mechanics of critical phenomena. It has already resulted in the attribution of two Fields Medals, in 2006 to Wendelin Werner for his work with Greg Lawler and Oded Schramm  on Brownian intersection exponents \cite{LSW01a,LSW01b,LSW02} and the Mandelbrot conjecture \cite{LSWMand}, and in 2010 to Stanislav Smirnov for his proof of the convergence on the hexagonal lattice of site percolation interfaces to SLE$_6$ \cite{Smir01}, and Ising model interfaces to SLE$_3$ \cite{Smir10,CS12}. Interestingly, while planar self-avoiding walks are strongly expected to converge in the continuum to SLE$_{8/3}$ \cite{LSW2004}, a proof has so far eluded even the best mathematical minds. Notice, however, that Nienhuis' 1982 famous prediction that the SAW connective constant equals $\mu=\sqrt{2+\sqrt{2}}$ on the hexagonal lattice was proven 30 years later  \cite{DumSmi12}! (See also \cite{BBDDG14} for the corresponding SAW critical adsorption fugacity $1+\sqrt{2}$).

\subsection{Higher polymer topologies} 
\subsubsection{Stars.}Outside the usual linear chain case, other \emph{polymer topologies} are possible, the most natural one being that of  a star $\mathcal S_L$, made up of $L$ self- and mutually avoiding arms of similar or equal lengths $N$. While the typical size of a star still scales like $N^\nu$, the star partition function or number of configurations (up to translations) now behaves for $N\to +\infty$ as,
\begin{equation}\label{eq:ZL}
\mathcal Z\left({\mathcal S_L}\right)\propto {\mu}^{LN} N^{\gamma_{\mathcal S_L}-1},
\end{equation}
where $\mu$ is the same connective constant as in the linear chain case \eqref{eq:gamma} \cite{WGLW86}, whereas $\gamma_{\mathcal S_L}$ is a critical exponent characteristic of the star topology, directly linked to the presence of an $L$-leg vertex. The single-chain case corresponds to $L = 1$, and $\gamma_{\mathcal S_1}=\gamma$. For $L = 2$,  
one has a two-leg star which is still a single linear polymer, hence  $\gamma_{\mathcal S_2}=\gamma$. 
For $L \geq 3$, the $\gamma_{\mathcal S_L}$'s  constitute a new set of independent critical exponents. In \emph{two-dimensions} for instance, their values were predicted via conformal invariance methods and Coulomb-gas methods \cite{BD86,S86} to be
\begin{equation}\label{gammaL}
\gamma_{\mathcal S_L}-1=\frac{1}{64} \big(4+9L(3-L)\big),
\end{equation}
in good agreement with the numerical results of exact enumeration and Monte-Calo techniques \cite{LWWMG85,WGLW86}. For $L=1,2$, one recovers $\gamma$ as in \eqref{nugamma}. 

Another illustrative example is that of a uniform star at the $\Theta$-point in three dimensions, which is the upper-tricritical dimension, where only \emph{logarithmic} corrections to  Brownian or random walk behavior occur. Its partition function can be shown via a direct renormalization of the relevant Edwards model to scale as \cite{BD88},
\begin{equation*}
\mathcal Z^{\Theta}\left({\mathcal S_L}\right)\propto {\mu_{\Theta}}^{LN} (\log N)^{-\frac{1}{22}{{L}\choose{3}}}, 
\end{equation*}
where $\mu_{\Theta}$ is the (model-dependent) connective constant at the $\Theta$-point, and where the log-power is expected to be universal.

Beyond that of stars and combs \cite{Gaunt86,LWWMG85,WGLW86}, the most general case of self-avoiding or $\Theta$-point polymer networks of arbitrary but fixed topologies was considered in any dimension via an extensive treatment of their configurational critical exponents, and published thirty years ago in the same {\it Journal of Statistical Physics}  \cite{D89}. It elaborated on the results of earlier Letters, the first one covering the bulk case in a good solvent \cite{BD86}, motivated by the results of \cite{S86} in two-dimensions, and followed by extensions to  networks at the ordinary surface transition, or at the  $\Theta$-point \cite{DS86,DS87,BD88}. The present article %, \textcolor{blue}{besides our recent work \cite{DG2019},} 
 can be seen as a distant-in-time sequel to these works (see also \cite{OB88}).
\subsubsection{Boundaries and bridges.}
\begin{figure}[h!]
\centering
\includegraphics[angle=0,scale =0.72] {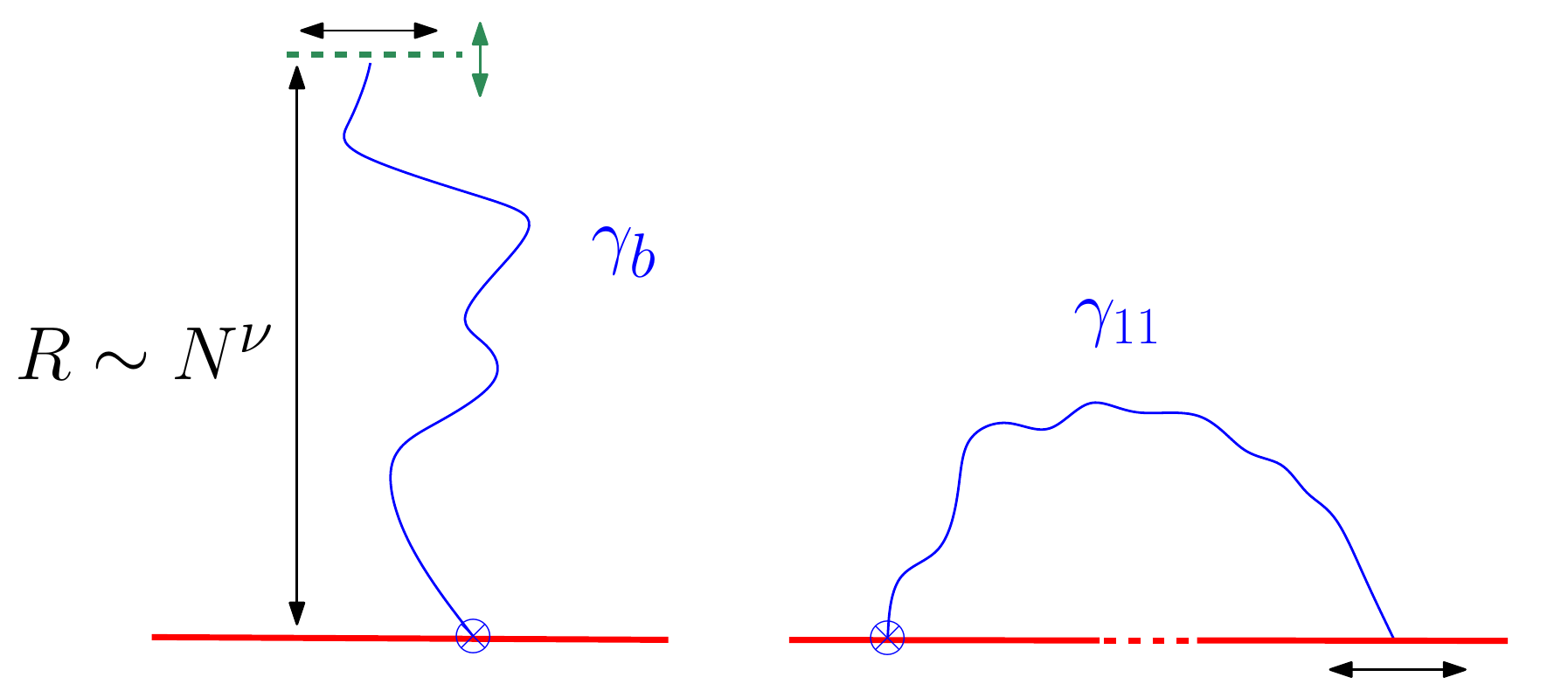}
 \caption{A polymer chain of length $N$ anchored at a surface, in bridge configurations below an hyperplane, with entropic exponent $\gamma_b$ (left), and in surface arch configurations with entropic exponent $\gamma_{11}$ (right). In each case, and to eliminate overall translational invariance, one (and only one) vertex ($\otimes$) is fixed along the surface. Compared to the motion of the second surface vertex in the arch configuration, the top-most vertex of the bridge configuration is free to move along a further direction perpendicular to the surface, by distances of order $R\sim N^{\nu}$, resulting in the simple scaling law $\gamma_b=\gamma_{11}+\nu$.}
 \label{fig:bridge1bis}
\end{figure}
 We recently revisited this theory to show that it can be extended to describe the  statistics of self-avoiding walks in \emph{bridge} or  \emph{worm} configurations \cite{DG2019}. As a result, the bridge configurational exponent was shown to equal 
\begin{equation}\label{eq:gbnug11bis}\gamma_b=\gamma_{11}+\nu,
\end{equation}
 where $\gamma_{11}$ is the usual configurational arch surface exponent at the ordinary surface transition corresponding to Dirichlet boundary conditions. The simple geometrical argument for this identity is recalled in Fig. \ref{fig:bridge1bis}.  In the presence of increasingly attractive surface interactions, a {\it special surface phase transition} may occur, where the chain contracts towards the surface, just before collapsing onto it.  At this special transition point, the bridge exponent is
\begin{equation}\label{eq:gbspg11bis}
\gamma_b({\rm sp})=\frac{1}{2}\left[\gamma_{11}({\rm sp})+\gamma_{11}\right] + \nu,
\end{equation}
where $\gamma_{11}({\rm sp})$ is now the special transition arch surface exponent \cite{DG2019}. 
 The worm configurational exponent was found to obey $\gamma=\gamma_w +\nu$, where $\gamma$ is the usual entropic exponent of a bulk SAW. The shift by $\nu$ here, as well as in Eqs. \ref{eq:gbnug11bis} and \ref{eq:gbspg11bis}, is brought about by the same mechanism as that described in Fig. \ref{fig:bridge1bis}. 
These results were based on the general theory yielding the configurational critical exponents associated with polymer networks of any topology, to which we now turn.
\subsubsection{General configurational exponents.} 
\begin{figure}[h!]
\centering
\includegraphics[angle=0,scale =0.52] {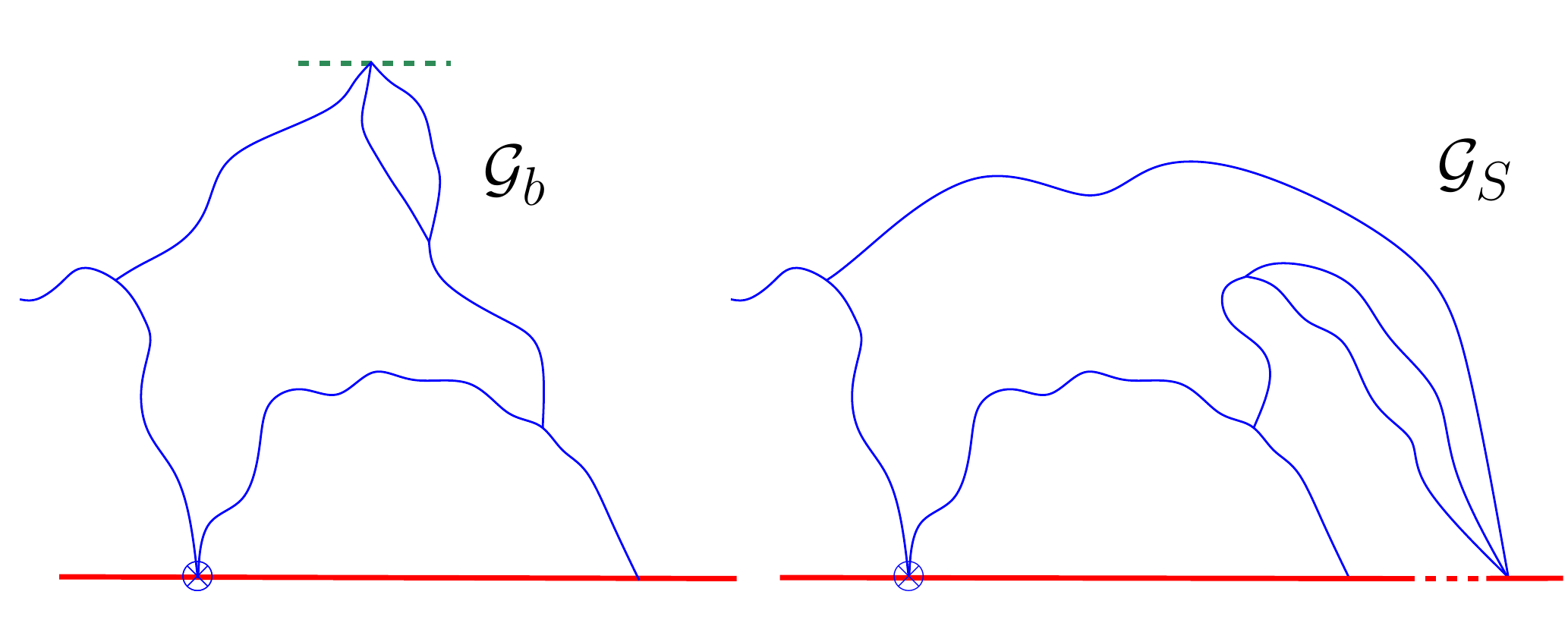}
 \caption{Polymer networks interacting with a surface. To eliminate overall translational invariance, one (and only one) vertex ($\otimes$) is fixed along the surface. Both networks have $\mathcal N=8$  lines, with respective lengths all scaling like $N$, with $N \to \infty$. Both networks possess 4 bulk vertices, of which $n_1=1$ has one line, and $n_3=3$ have three lines attached. Network ${\mathcal G}_S$ on the right has 3 surface vertices, $n_1^S=1$ with one leg, $n_2^S=1$  with two legs, and $n_3^S=1$  with three legs. Network ${\mathcal G}_b$ on the left is in a \emph{bridge} configuration. It has the same surface vertices, except for the 3-leg \emph{surface} vertex being replaced by the (top-most) 3-leg \emph{bridge} vertex.}
 \label{fig:network3}
\end{figure}
Consider surface-attached polymer networks such as those represented in Fig. \ref{fig:network3}. One arbitrary boundary vertex is fixed to eliminate overall translations.   
The partition function or number of configurations of such a self-avoiding (monodisperse) $d$-dimensional polymer network ${\mathcal G_S}$, made of $\mathcal N$ chains of equal lengths $N$, is given for $N\to +\infty$ by an expression similar to \eqref{eq:gamma} and \eqref{eq:ZL},
\begin{equation}\label{ZG}
{\mathcal Z}\left({\mathcal G_S}\right)\propto {\mu}^{\mathcal N N} N^{\gamma_{\mathcal G_S}-1},
\end{equation}
with the same the growth constant $\mu$  as in the single chain case. The configurational critical exponent $\gamma_{\mathcal G_S}$ is given by the explicit exact formula \cite{BD86,DS86,D89,OB88}, 
\begin{equation} \label{eq:key}
\gamma_{\mathcal G_S} = \nu \left[d{\mathcal V}+(d-1)({\mathcal V}_S-1) - \sum_{L \ge 1} (n_Lx_L+n_L^Sx_L^S)\right]-({\mathcal N}-1).
\end{equation}
Here ${\mathcal V}$ is the number of bulk vertices, ${\mathcal V}_S$ is the number of surface vertices, $n_L$ is the number of $L$-leg vertices floating in the bulk, while $n_L^S$ is the number of $L$-leg vertices on or constrained close to, the surface. ${\mathcal N}$ denotes the number of chains in the network.  
The intuitive meaning of formula \eqref{eq:key} for such a configurational  exponent $\gamma_{\mathcal G_S}$ is clear: the first two terms correspond, {via the correlation lengh exponent $\nu$,} to the Euclidean phase space of the (bulk and surface) vertices of the network, the $x_L$ and $x_L^S$ (universal) critical exponents correspond to the entropic reduction in phase space induced by both linkage and self- or mutual-avoidance in each $L$-star vertex, and the last term, $\mathcal N-1$, corresponds to the constraint of \emph{monodispersity} in the $\mathcal N$ arms of the network ({\it i.e.}, their respective lengths, or monomer numbers,  all scale like $N$, with $N\to +\infty$).

For an unconstrained network $\mathcal G$ in the \emph{bulk}, with no surface attached vertices, $d{\mathcal V}$ must be replaced by $d({\mathcal V}-1),$ as otherwise one is counting all translates, and the formula simply becomes \cite{BD86,D89}
 \begin{equation} \label{eq:keybis}
\gamma_{\mathcal G} = \nu \left[d({\mathcal V}-1) - \sum_{L \ge 1} n_Lx_L\right]-({\mathcal N}-1).
\end{equation}
For an illustrative example, let us return to the star bulk case $\mathcal G:=\mathcal S_L$, for which $\mathcal V=L+1$, $\mathcal N=L$, $n_1=L$, and $n_L=1$. Eq. \eqref{eq:keybis} then gives $$\gamma_{\mathcal \mathcal S_L} =\nu (Ld-Lx_1-x_L)-(L-1).$$ For $L=1,2$, we have
\begin{align}
\gamma_{\mathcal S_1}=\nu(d-2x_1);\quad \gamma_{\mathcal S_2}=\nu (2d-2x_1-x_2)-1.
\end{align}
Since $\gamma_{\mathcal S_1}=\gamma_{\mathcal S_2}=\gamma$, we have $\nu (d-x_2)=1$, hence the known identity $x_2=d-1/\nu$ \cite{Nienhuis87,D89}.

 A non-trivial result, of course, is that such a reduction in \eqref{eq:key} or \eqref{eq:keybis} to \emph{individual vertices} holds true;  it can be obtained in 2 dimensions from \emph{conformal field theory} \cite{BD86,DS86,D89}, or from a \emph{two-dimensional quantum gravity} approach \cite{DK88,DK90}. In generic dimension $d$, one resorts to the \emph{multiplicative renormalization theory} of polymer networks \cite{D89,SFLD92}, which is an extension of that of polymer chains \cite{JdC80,JdC81,JdC89,LS99}.  

The values of the universal critical exponents $\nu$, $x_L$ and $x^{S}_L$ naturally all depend on the space dimension and on the universality class of the polymer problem considered: SAWs, polymer chains at the $\Theta$-point, or mutually-avoiding random walks, and, concerning the surface case, ordinary or special transition boundary conditions. All these values will be made precise below. One of the aims of this work is to present these exponents in the two-dimensional case, often initially derived in theoretical physics from Coulomb gas, conformal invariance or Bethe Ansatz methods, from a  more unified perspective, and in particular  to place them in broader families of critical exponents, that can be derived within the recent SLE and Liouville Quantum Gravity mathematical frameworks \cite{D04,DMS2014,DuSh2011}.

Note that the simplest case where the multiplicative renormalization theory presented above can be established in a straightforward way is that of {\it Brownian networks} made up of simple random walks, or, in the continuum, Brownian chains.  In this case, the scaling exponent $\nu=1/2$, and the bulk and Dirichlet surface $L$-leg exponents are (see, e.g., \cite{BDCMP87})
\begin{equation}\label{eq:Brown}
x^{\rm B}_L=\frac{L}{2}(d-2),\,\,\,\,x^{S,{\rm B}}_L=\frac{L}{2}d.
\end{equation}
Note their proportionality to $L$, coming from the statistical independence of Brownian paths. In this case, formulae \eqref{eq:key} and \eqref{eq:keybis} can be reduced to simpler expressions, using the topological identities for the total chain number $\mathcal N$,  independent loop number $\mathscr L$, and bulk and surface vertex numbers $\mathcal V$ and ${\mathcal V}_S$,
\begin{eqnarray}
\mathcal N&=&\frac{1}{2}\sum_{L\geq 1}L (n_L+n_L^S), \,\,\,\,\mathscr L=\frac{1}{2}\sum_{L\geq 1} (L-2) (n_L+n_L^S) +1,\\
\mathcal V&=&\sum_{L\geq 1}n_L,\,\,\,\,{\mathcal V}_S=\sum_{L\geq 1}n_L^S.
\end{eqnarray}
For a pure \emph{bulk}  Brownian network, this together with \eqref{eq:Brown} gives for Eq. \eqref{eq:keybis}
\BE\label{eq:topo}
\gamma^{\rm B}_{\mathcal  G}-1=-{\mathscr L}d/2,
\EE
a well-known result in the computation of polymer Feynman diagrams (see, e.g., \cite{BDCMP87,D89}). When the Brownian network is attached onto a Dirichlet surface, one finds for \eqref{eq:key} \cite[Section 6]{D89}
\BE\label{eq:topo2}
\gamma^{\rm B}_{\mathcal  G_S}-1=-{\mathscr L}d/2 -\frac{1}{2} ({\mathcal V}_S-1)-\frac{1}{2} {\mathcal L}_S,
\EE
where ${\mathcal L}_S:=\sum_{L\geq 1}Ln_L^S$ is the number of chains terminally touching the surface. 
\section{Polymer networks and bridges}
 \subsection{Bridges }
\begin{figure}
\centering
\includegraphics[angle=0,scale =0.52] {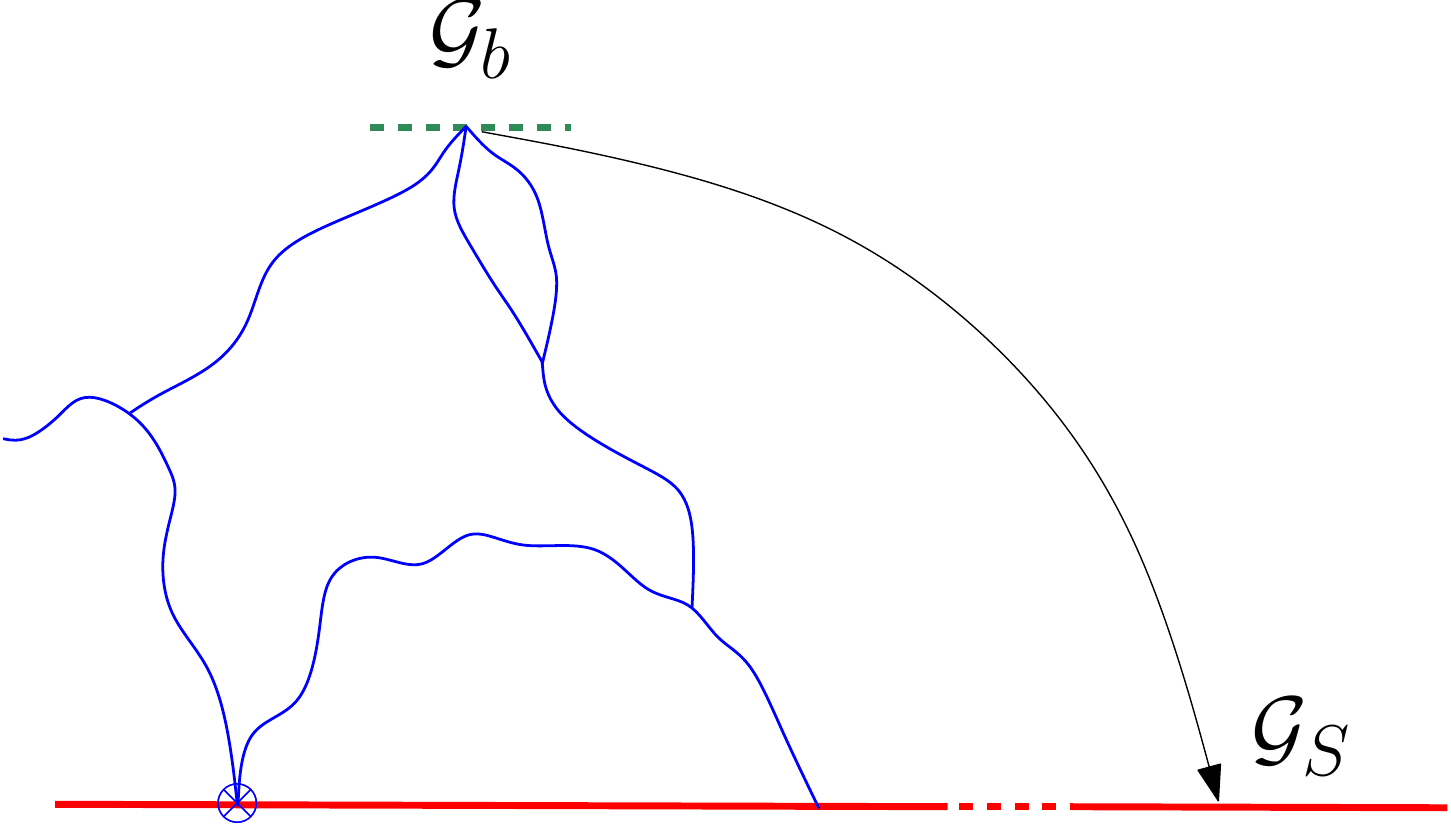}
 \caption{Transformation of the bridge polymer network ${\mathcal G}_b$ into a surface network ${\mathcal G}_S$ in Fig. \ref{fig:network3}. The single extra spatial degree of freedom of the movable hyperplane of the (3-leg) bridge vertex with respect to the corresponding (3-leg) surface vertex induces the simple scaling relation of the respective configuration exponents, $\gamma_{{\mathcal G}_b}=\gamma_{{\mathcal G}_S}+\nu$.}
 \label{fig:network5}
\end{figure}

 A {\it bridge} configuration $\mathcal G_b$ in a polymer network anchored to a surface appear when a given $L$-leg vertex is the top-most point of the network (Fig. \ref{fig:network3}, left). This constraint is equivalent to the appearance of a movable virtual top-most hyperplane, contributing in Eq. \eqref{eq:key} a boundary exponent $x_L^{S}$,  as in the right-most polymer network $\mathcal G_S$ in Fig. \ref{fig:network3} that results from the topological move as represented in Fig. \ref{fig:network5}. However, because the virtual top-most hyperplane is movable, the said top-most $L$-vertex in $\mathcal G_b$ contributes one unit to the number ${\mathcal V}$ of bulk vertices in Eq. \eqref{eq:key}, whereas the boundary vertex in $\mathcal G_S$ contributes one unit to the number  ${\mathcal V}_S$ of boundary vertices in Eq. \eqref{eq:key}. We thus have in all generality for the pair of networks in Figs. \ref{fig:network3}, \ref{fig:network5},
  \begin{equation}\label{eq:gbnugS}\gamma_{\mathcal G_b}=\gamma_{\mathcal G_S}+\nu,
\end{equation}
which yields a direct generalisation of identity \eqref{eq:gbspg11bis}. 

\subsection{Multi-bridges}\label{multibridge}
\begin{figure}[h!]
\centering
\includegraphics[angle=0,scale =0.52] {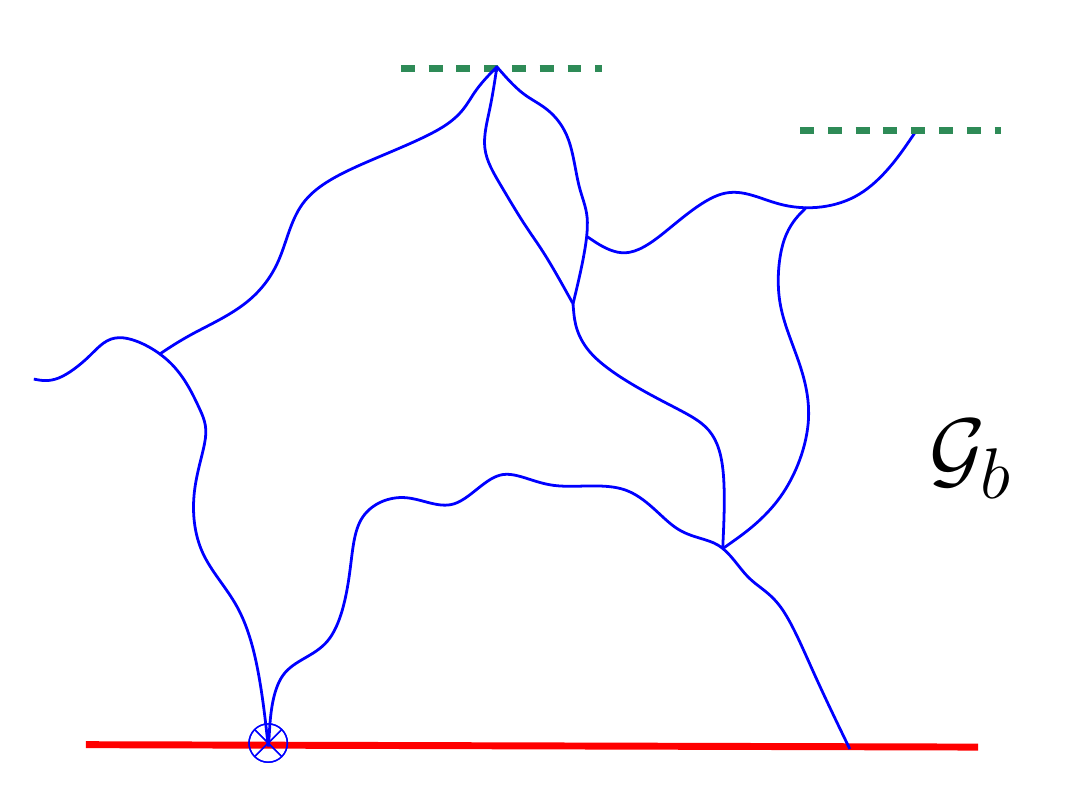}
 \caption{A \emph{multi-bridge} polymer network ${\mathcal G}_b$ anchored at a surface. To eliminate overall translational invariance, one (and only one) vertex ($\otimes$) is fixed along the surface. This network has $n_1^b=1$ single bridge vertex and $n_3^b=1$ 3-leg bridge vertex, where the polymer strands are \emph{locally confined} below their respective hyperplanes ({\it e.g.}, up to the next connected vertices). The lengths of the $\mathcal N=11$ chains of the network all scale like $N$, with $N \to \infty$.}
 \label{fig:network4}
\end{figure}
The bridge theory presented so far can be extended to \emph{multi-bridge} network configurations. These are defined only \emph{locally}: a given $L$-vertex  is said to be an $L$-multi-bridge if each of the $L$ chains attached to it stay below (or above) the vertex at least up to the second vertex to which it is bound. An example is given in Fig. \ref{fig:network4}.  Such a multi-bridge vertex contributes 
a term $-\nu x_L^{S}$, together with a free-volume term $+\nu d$, to the entropic exponent $\gamma_{\mathcal G}$. Denoting by $n_L^{b}$ the numbers of $L$-bridge vertices, we therefore arrive at a more general form of the entropic exponent of a multi-bridge network $\mathcal G_b$,
\begin{equation} \label{eq:keyter}
\gamma_{\mathcal G_b} = \nu \left[d{\mathcal V}+(d-1)({\mathcal V}_S-1) - \sum_{L \ge 1} \big(n_Lx_L+(n_L^S+n_L^b)x_L^S\big)\right]-({\mathcal N}-1),
\end{equation}
  where now the bulk and surface vertex numbers are, respectively,
  \BE \label{Vcount}
  {\mathcal V}=\sum_{L\geq 1}(n_L+n_L^b),\,\,\,\,\,\,{\mathcal V}_S=\sum_{L\geq 1}n_L^S.
  \EE
  \subsection{Special transition}
  If there is an \emph{attractive} surface fugacity $a=\exp(-\epsilon/k_BT),$ where $\epsilon$ is the energy associated with a monomer of the walk lying in the surface, then this has no effect on the critical point or critical exponent of the network provided that $a$ is less than its critical value $a_c > 1,$ where $a_c$ is called the {\em critical fugacity}. Precisely at the critical fugacity, the exponent changes discontinuously. This is called the {\em special transition}, and the corresponding expression to Eq. (\ref{eq:key}) follows straightforwardly by replacing the Dirichlet scaling dimensions $x_L^S$  by their values at the special transition, $x_L^S({\rm sp})$. 
This extension to the case of polymer networks at the special transition in two-dimensions was given in \cite{BDO98}, who also included the \emph{mixed} case of some surface vertices being at the critical fugacity and others not. In this mixed case, the general formula \eqref{eq:key}, as well as the multi-bridge one \eqref{eq:keyter} obviously generalize to 
\begin{equation} \label{eq:keysp}
\gamma_{\mathcal G_S} = \nu \left[d{\mathcal V}+(d-1)({\mathcal V}_S-1) - \sum_{L \ge 1} \big(n_Lx_L+n_L^Sx_L^S+n_L^S({\rm sp})x_L^S({\rm sp})\big)\right]-({\mathcal N}-1),
\end{equation}
\begin{equation} \label{eq:keytersp}
\gamma_{\mathcal G_b} = \nu \left[d{\mathcal V}+(d-1)({\mathcal V}_S-1) - \sum_{L \ge 1} \big(n_Lx_L+(n_L^S+n_L^b)x_L^S+n_L^S({\rm sp})x_L^S({\rm sp})\big)\right]-({\mathcal N}-1),
\end{equation}
where $n_L^S({\rm sp})$ represents the number of surface $L$-leg vertices at the special transition point, and where now $\mathcal V=\sum_{L\geq 1}\left(n_L+n_L^b\right)$ and $\mathcal V_S=\sum_{L\geq 1}\left(n_L^S+n_L^S({\rm sp})\right)$. 
 
For $a > a_c$ the location of the critical point varies monotonically with $a,$ and the exponents change to integers, corresponding to poles in the generating function. In this paper we will only consider the situation $a \le a_c,$ which corresponds to the most interesting physics.
\subsection{A consistency check}\label{Consistency}
\begin{figure}[h!]
\centering
\includegraphics[angle=0,scale =1.72] {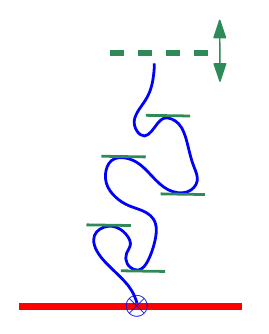}
 \caption{Consistency check of the scaling theory of local bridges: Any local extremum along an anchored polymer chain (which may be part of a larger network $\mathcal G_S$), can be seen as a {\sl virtual two-leg bridge vertex}, yielding a reduced-entropy contribution  $-\nu x_2^{S}=-\nu d$ to $\gamma_{\mathcal G}$ in Eq. \eqref{eq:key}, together with a volume contribution $+\nu d$ due to its bulk degrees of freedom, in effect yielding \emph{no} overall contribution. The same is true in $\Theta$-point conditions where $x_2^{S,\Theta}=d$ also holds.}
 \label{fig:2top}
\end{figure}
A peculiar case in the network-multibridge theory presented here is that of local extremal points along the chains making up the network (Fig. \ref{fig:2top}). Even though they fleetingly appear in the configuration space, they can be considered as genuine two-leg vertices associated with a local bridge point ({\it i.e.},  a local top-most or bottom-most point). Each then should contribute a $-\nu x_2^{S}$ Dirichlet surface term, together with a usual volume term $+\nu d$ to configuration exponent $\gamma_{\mathcal G_S}$ \eqref{eq:keybis}.  However, 
it is a fundamental fact \cite[Section 6.5.1]{D89} (see also \cite{DDE83}) that in any dimension $d$, the two-leg \emph{ordinary transition} surface exponent for a polymer chain with excluded volume takes its \emph{Brownian} or random walk value \eqref{eq:Brown} 
\BE\label{x2S}
x_2^S=x_2^{S,\mathrm{B}}=d.
\EE  This also holds for polymer chains at the $\Theta$-point, as we shall see below. Thus, the two contributions exactly cancel, and extremal points along a chain yield no contribution to the entropy of the chain, as it must. Note that this argument can be reversed to yield the simplest proof that one necessarily has,  for Dirichlet ordinary boundary conditions, the 2-leg surface exponent $x_2^S\equiv d$, irrespective of the universality class of the polymer chain (SAW, $\Theta$-point, Brownian).

\section{Multiple-path critical exponents}
In this section, we first briefly recall the values of bulk exponents $x_L$, surface exponents $x_L^S$, and special surface exponents $x_L^S(\mathrm{sp})$, as well as some generalisations of them. In a second step, in the two-dimensional case, we provide a direct, general technique \cite{D04,DMS2014} to compute them with the help of Liouville quantum gravity (LQG) and of the famous Knizhnik-Polyakov-Zamolodchikov (KPZ) relation \cite{KPZ88}, adapted to the Schramm-Loewner Evolution (SLE) \cite{Schramm2000}. This will in particular be useful in the case of polymer networks at the $\Theta$-point in two-dimensions. This approach is naturally related to multiple SLE theory \cite{BBK05,BPW18,Dubedat2007,Gra07,KL07,KP16,Law09a,HWu18,WuZhan17} (For a recent mathematical perspective, see \cite{Peltola19,PeltWu19}.)
\subsection{A short survey} 
In two-dimensions, the bulk conformal weights $(\frac{1}{2})x_L,$ and surface conformal weights $x_L^S,$  associated with $L-$vertices, are explicitly given by 
\begin{equation} \label{eq:xnorm}
x_L=\frac{1}{48}(3L-2)(3L+2), \,\,\,\,\,\,  x_L^S=\frac{1}{8}L(3L+2),
\end{equation}
with $1/\nu=2-x_2=4/3$. (See Refs. \cite{Nienhuis82,Nienhuis87} for the $L=1,2$ bulk cases, Ref. \cite{Cardy84} for the $L=1$ boundary case, and \cite{S86,S87,DK88,BB89,D04} for the general bulk case, and \cite{DS86,GB89,BEG89, BE94,BS93,EE93,D04} for the general boundary case). In $d$-dimensions, they have the general form \cite{D89},
%$$x_L=\frac{L}{2}(d-2)+x'_L,\,\,\,\,\,\, x^S_L=\frac{L}{2}d+x'^S_L,$$
$$x_L=x^{\rm B}_L+x'_L,\,\,\,\,\,\, x^S_L=x^{S,{\rm B}}_L+x'^S_L,$$
where the first terms are the Brownian scaling dimensions \eqref{eq:Brown}, whereas the second ones, $x'_L,\, x'^S_L$, represent the anomalous contributions from self- and mutual avoidance. At first orders in $\varepsilon:=4-d$, the latter are \cite{JdC89,D89}
$$x'_L=\frac{\varepsilon}{8}L(L-1)+\left(\frac{\varepsilon}{8}\right)^2\frac{L}{4}\left(-8L^2+33L-23\right)+O(\varepsilon^3),$$ and 
\BE\label{eq:anom}x'^S_L=\frac{\varepsilon}{8}L(L-2)+O(\varepsilon^2),
\EE
 while being also explicitly known to next order in $\varepsilon$ \cite{Dieh86,DD80,DD81,EE93,FH97,SFLD92}. For $d\geq 4$, $x'_L=0,x'^S_L=0$, with known logarithmic corrections to Brownian network partition functions at  $d=4$ \cite{D89}. 

As we have seen, Eqs. \eqref{eq:key},  \eqref{eq:keybis}, and  \eqref{eq:keyter} naturally hold in the case of random walks or Brownian chains, for which $\nu=1/2$ and the scaling exponents take the above mentioned  Brownian values \eqref{eq:Brown}. They also hold  in the mixed case of a network made of \emph{mutually-avoiding} (M) random walks or Brownian chains, for which the bulk and surface scaling exponents are in two-dimensions \cite{DKw88,D98,LSW01a,LSW01b} 
\BE\label{eq:MAW}
x_L^{\rm M}=\frac{1}{12}(4L^2-1),\,\,\, x_L^{\rm M,S}=\frac{1}{3}L(2L+1).
\EE
In $d=4-\varepsilon$ dimensions the bulk exponent is \cite{BDCMP87}
$$x_L^{\rm M}=x_L^{\rm B}+\frac{\varepsilon}{4}L(L-1)-\left(\frac{\varepsilon}{4}\right)^2L(L-1)(2L-5)+O(\varepsilon^3),$$
while for $d\geq 4$, $x_L^{\rm{M}}=x_L^{\rm B}$ and $x_L^{\rm M,S}=x_L^{S,\rm B}$, with known logarithmic corrections at $d=4$  \cite{BDCMP87}.

Notice that for SAWs, as announced in Section \ref{Consistency}, the 2-leg Dirichlet surface exponent in two-dimensions is from \eqref{eq:xnorm} $x_2^S=2=d=x_2^{S,\mathrm{B}}$, i.e., the Brownian one, in agreement with Eq. \eqref{x2S}, and that in $d$-dimensions, the anomalous addition $x'^S_L$ \eqref{eq:anom} to the Brownian surface exponent vanishes for $L=2$ at first-order (as it does to all orders). 

At the special transition point, the scaling dimensions $x_L^S$ \eqref{eq:xnorm} are replaced by their values in two-dimensions, obtained in Refs. \cite{FS94,BY95,BY95bis,YungBat95} (see \cite{GB89} for the $L=1$ case and see also \cite{BEG89,BE94,BD89}),
\begin{equation}\label{eq:xsurfsp}
x_L^S({\rm sp})=\frac{3}{8}L^2-\frac{3}{4}L+\frac{1}{3}.
\end{equation}
%This extension to the case of polymer networks at the special transition in two-dimensions was given in \cite{BDO98}, who also included the \emph{mixed} case of some surface vertices being at the critical fugacity and others not. In this mixed case, the general formula \eqref{eq:key}, as well as the the multi-bridge one \eqref{eq:keyter} obviously generalize to 
%\begin{equation} \label{eq:keysp}
%\gamma_{\mathcal G_S} = \nu \left[d{\mathcal V}+(d-1)({\mathcal V}_S-1) - \sum_{L \ge 1} \big(n_Lx_L+n_L^Sx_L^S+n_L^S({\rm sp})x_L^S({\rm sp})\big)\right]-({\mathcal N}-1),
%\end{equation}
%\begin{equation} \label{eq:keytersp}
%\gamma_{\mathcal G_b} = \nu \left[d{\mathcal V}+(d-1)({\mathcal V}_S-1) - \sum_{L \ge 1} \big(n_Lx_L+(n_L^S+n_L^b)x_L^S+n_L^S({\rm sp})x_L^S({\rm sp})\big)\right]-({\mathcal N}-1),
%\end{equation}
%where $n_L^S({\rm sp})$ represents the number of surface $L$-leg vertices at the special transition point, and where now $\mathcal V=\sum_{L\geq 1}\left(n_L+n_L^b\right)$ and $\mathcal V_S=\sum_{L\geq 1}\left(n_L^S+n_L^S({\rm sp})\right)$. 
 %For $a > a_c$ the location of the critical point varies monotonically with $a,$ and the exponents change to integers, corresponding to poles in the generating function. In this paper we will only consider the situation $a \le a_c,$ which corresponds to the most interesting physics.
\subsection{LQG-KPZ derivation of 2D exponents}\label{DNLQG}
\subsubsection{Introduction.}
In \cite{D04,BDLH05}, a systematic way is provided to compute critical exponents of SLE (i.e., of the $O(n)$ or Potts models in two-dimensions), by extensively using a technique imported from {\sl Liouville quantum gravity}. It consists in calculating similar ``quantum'' exponents as measured with the LQG random measure  (which represents the random area measure generated in a 2D critical random lattice). Then, Euclidean critical exponents, as defined with respect to the usual Lebesgue measure in the plane, are expressed in a unique way in terms of the former quantum ones via the celebrated KPZ relation \cite{KPZ88,KPZbis,DKaw89}. The interest of the method is that some quantum exponents can be calculated using random matrix theory (RMT), as was originally done in the late '80s \cite{Kaz86,DK88,DK90}. Furthermore, as shown in \cite{She16,DMS2014}, they enjoy fundamental {\bf conformal welding} properties (by ``cutting and gluing''), which    
make their computation remarkably simple, and this without recourse to RMT nor Coulomb gas methods. The KPZ relation has been mathematically proven in various settings in the last decade \cite{DuSh2009,DuSh2011,RhVa11,DRSV2014,BD2010,BJRV13}, and, as we shall se below,  the quantum gravity techniques advocated here \cite{BD2000,BD2003,D04,BDLH05} are part and parcel of  the conformal welding of so-called quantum wedges and cones, introduced in the  recent mathematical theory of Liouville quantum gravity as developed in \cite{DMS2014} (see also \cite{BDICM2014,GHS2019} for introductions).\\ 

The Schramm-Loewner Evolution, SLE$_\kappa$, is a canonical conformally invariant process in the plane, either in the disk  $\mathbb D$ (called radial), or in the half-plane $\mathbb H$ (called chordal) or in the whole-plane $\mathbb C$. It crucially  depends on a single real parameter $\kappa \geq 0$.  For $0\leq \kappa\leq 4$, the random paths generated are \emph{simple}, whereas for $4<\kappa<\infty$ they continually \emph{self-intersect} (but do not self-cross). For $\kappa \geq 8$, they are Peano curves in the plane. The relation to the critical $O(n),   n\in[0,2]$, model  in the plane is given by $n=-2\cos\left(4\pi/\kappa\right)$, with $\kappa\in [8/3,4]$ for the \emph{dilute} critical point, and $\kappa\in [4,8]$ for the so-called \emph{dense} phase of the $O(n)$-model \cite{BD2003,D04,KN2004}. For the  Potts critical model with parameter $Q\in [0,4]$, there is complete equivalence to the universality class of the $O(n)$ dense phase, via the simple identity $n=\sqrt{Q}$ \cite{Nienhuis82,Nienhuis87,BD87}. 
\subsubsection{KPZ relation and duality.}
In Refs. \cite[Section 11.1]{D04} and \cite[Section 10.2]{BDLH05}, the following KPZ formulae, adapted to computations with SLE$_\kappa$, were proposed,
\begin{eqnarray}\label{eq:KPZ}
\mathcal U_\kappa(\Delta)&=&\frac{1}{4}\Delta \left(\kappa \Delta+4-\kappa\right),\\ \label{eq:BtoBfirst}
\mathcal V_\kappa(\Delta)&=&\mathcal U_\kappa\left[\frac{1}{2}\left(\Delta +1-\frac{4}{\kappa}\right)\right]
=\frac{1}{16\kappa}\left[\kappa^2 \Delta^2-(4-\kappa)^2\right].
\end{eqnarray}
Function $\mathcal U_\kappa$ transforms SLE$_\kappa$ quantum boundary scaling exponents $\Delta$  into their Euclidean boundary counterparts, $x^S=\mathcal U_\kappa(\Delta)$, whereas function $\mathcal V_\kappa$ transforms these into their Euclidean \emph{bulk} counterparts, $x=2\mathcal V_\kappa(\Delta)$ \cite{D04,BDLH05}. As explained in detail there, these were precisely devised to be insensitive to the usual change of parameterization of Liouville quantum gravity between the $\kappa\in[0,4]$ and $\kappa\in(4,\infty)$ ranges.  Actually,  for $\kappa\in [0,4]$, Eqs. \eqref{eq:KPZ} is the usual KPZ relation \cite{DuSh2011} $U_\gamma(\Delta)=\frac{1}{4}\Delta \left(\gamma^2\Delta+4-\gamma^2\right)$, where $\gamma=\sqrt{\kappa}\in[0,2]$ is the Liouville measure parameter.  When $\kappa\in (4,\infty)$, it is the \emph{dual} KPZ relation \cite{D04,BDLH05,BD2010,BJRV13,BDICM2014} $U_{4/\gamma}(\Delta)$, where $\gamma=\sqrt{16/\kappa}\in (0,2)$. 

\subsubsection{Multiple SLE exponents.} 
\begin{figure}[h!]
\centering
\includegraphics[angle=0,scale =1.52] {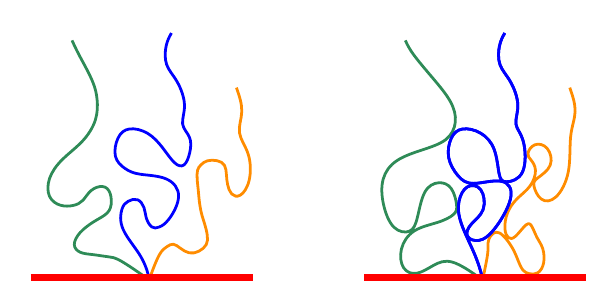}
 \caption{$L=3$ multiple chordal SLE$_\kappa$ paths rooted at a surface in two-dimensions. {\sl Left:} For $\kappa\leq 4$, the paths are simple and do not touch each other, nor the surface outside of the root. {\sl Right:}  For $\kappa > 4$, the chains are non-simple,  bounce repeatedly onto each other and onto the surface.}
 \label{fig:DN}
\end{figure}
%\begin{figure}[h!]
%\centering
%\includegraphics[angle=0,scale =1.52] {DN.pdf}
% \caption{$L=3$ multiple chordal SLE$_\kappa$ paths rooted at a surface in two-dimensions. {\sl Left:} Dirichlet case for $\kappa\leq 4$, the paths are simple and do not touch each other, nor the surface outside of the root, so that $x_L^{S,\kappa}=x_L^{S,\kappa}(\mathrm{ord})$. {\sl Right:}  Neumann case for $\kappa > 4$, the chains are non-simple,  bounce repeatedly onto each and onto the surface, so that $x_L^{S,\kappa}=x_L^{S,\kappa}(\mathrm{sp})$.}
 %\label{fig:DN0}
%\end{figure}
 In the case of \emph{multiple}  SLE$_\kappa$ paths originating at a boundary point in $\mathbb H$ (Fig \ref{fig:DN}), the quantum boundary exponent  $\Delta$ is built by simply adding the individual contributions $2/\kappa$ of each path separately \cite{D04,BDLH05}, namely, 
 \BE\label{eq:deltaL}
 \Delta=\Delta_L^{\kappa}:=2L/\kappa.
 \EE 
Note that for $\kappa \leq 4$, $\Delta_L^\kappa$ is a usual (additive) quantum boundary exponent in LQG, whereas for $\kappa >4$, it represents the \emph{dual} quantum boundary exponent that is additive.  (See \cite[Section 12.1]{D04} and \cite[Section 10.5]{BDLH05}.) This leads to \cite{D04,BDLH05,HWu18,WuZhan17}
\begin{eqnarray}\label{eq:KPZL}
x^{S,\kappa}_L=\mathcal U_\kappa(\Delta_L^{\kappa})&=&\frac{L}{2\kappa} \left(2L+4-\kappa\right),\\ \label{eq:BtoBL}
x_L^{\kappa}=2\mathcal V_\kappa(\Delta_L^{\kappa})&=&\frac{1}{8\kappa}\left[4L^2-(4-\kappa)^2\right].
\end{eqnarray}  
One must then recall that the boundary behaviour of chordal SLE$_\kappa$ strongly depends on its phase (Fig. \ref{fig:DN}): If $\kappa\leq 4$, the random paths do not intersect the boundary line.  If $\kappa>4$, the SLE$_\kappa$ paths continually touch and bounce onto the boundary.  
Nevertheless,  \eqref{eq:KPZL} corresponds in both cases to the \emph{ordinary} boundary phase for the corresponding $O(n)$-model. 
\subsubsection{Conditioning boundary intersections ($\kappa>4$).}\label{conditioning}
%\begin{figure}[h!]
%\centering
%\includegraphics[angle=0,scale =1.52] {multipletheta.pdf}
% \caption{$L=3$ multiple chordal SLE$_{\kappa>4}$ paths in two-dimensions. For $\kappa>4$, chains continually bounce onto each other and onto themselves.  The paths can be  subject to various boundary conditions and constraints. {\sl Left:} Usual Neumann boundary conditions where the chains also bounce repeatedly onto the surface, with exponent $x_L^{S,\kappa}=x_L^{S,\kappa}(\mathrm{sp})$; {\sl Middle:} Dirichlet boundary conditions (as marked by two arrows) where the chains are forbidden to touch the surface outside of the root, with exponent $x_{L,j=2}^{S,\kappa}=x_L^{S,\kappa}(\mathrm{ord})$; {\sl Right:} in addition to previous Dirichlet boundary conditions, the two leftmost chains are conditioned not to intersect, as marked by a third arrow, with exponent $x_{L,j=3}^{S,\kappa}$.}
% \label{fig:multipletheta}
%\end{figure}
\begin{figure}[h!]
\centering
\includegraphics[angle=0,scale =1.32]{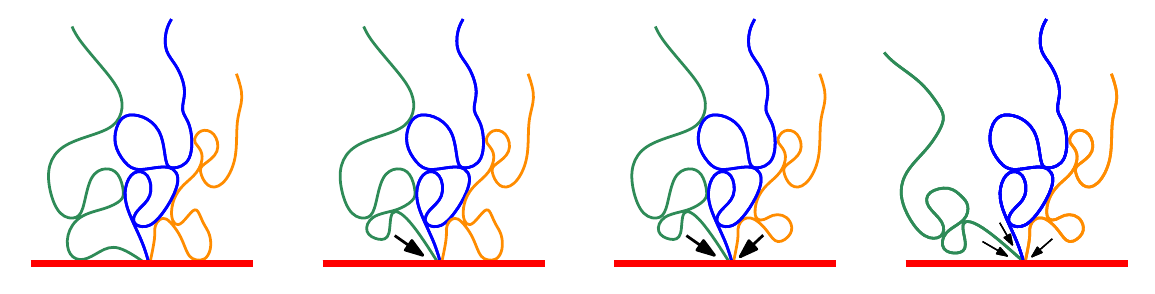}
 \caption{$L=3$ multiple chordal SLE$_{\kappa>4}$ paths in two-dimensions. For $\kappa>4$, they continually bounce onto themselves and onto each other.  The paths can be  conditioned in several ways. {\sl Left:} Ordinary boundary conditions where the paths repeatedly bounce onto the surface, with surface exponent $x_{L=3}^{S,\kappa}=x_{L=3,j=0}^{S,\kappa}$. {\sl Middle left:} Mixed boundary conditions (as marked by one arrow) where the leftmost path is conditioned not to hit the surface outside the root, with exponent $x_{L=3,j=1}^{S,\kappa}$. {\sl Middle right:} The leftmost and rightmost paths are conditioned not to intersect the boundary, as marked by two arrows, with exponent $x_{L=3,j=2}^{S,\kappa}$. {\sl Right:} Three non-intersection conditions are imposed, resulting in exponent  $x_{L=3,j=3}^{S,\kappa}$. Each arrow indicates the addition of a unit of quantum dimension \eqref{eq:U0+} in \eqref{eq:DeltaLj+}.}
 \label{fig:multiplethetabis}
\end{figure}
Let us introduce the positive inverse of function  $\mathcal U_\kappa$ \eqref{eq:KPZ},
\BE
 \label{eq:U-1}
\mathcal U^{-1}_\kappa(x)=\frac{1}{2\kappa}\left(\sqrt{16\kappa x+(4-\kappa)^2}+\kappa -4\right).
\EE
As shown in  \cite[Section 12.1]{D04}, and \cite[Section 10.5]{BDLH05},  for $\kappa>4$, thus for \emph{non-simple} SLE$_{\kappa >4}$ paths, one can condition the paths constituting an SLE$_\kappa$  $L$-star not to intersect the boundary, by inserting along the boundary `point operators'  with a non-vanishing quantum scaling dimension, 
\BE\label{eq:U0+}\mathcal U^{-1}_\kappa(0)=\vartheta(\kappa-4) \left(1-4/\kappa\right),
\EE
 where $\vartheta(x)=0, x<0; \vartheta(0)=1/2; \vartheta(x)=1, x>0$, is the Heaviside distribution.    
 For a total number $j\in \{0,1,2\}$ of such insertions, one obtains the first three boundary conditional cases depicted in Fig \ref{fig:multiplethetabis}. This operation can also be performed between any two paths, which, by conformal invariance, acts in the same way as an insertion in between the left-most path (or the right-most path) and the boundary (Fig. \ref{fig:multiplethetabis}, right-most case with $j=3$). Thus, one can take $0\leq j\leq L+1$ for the boundary case, and $0\leq j\leq L$ for the bulk case.  
 
 \emph{For $\kappa > 4$, {\bf dual} quantum boundary dimensions are additive in LQG}  (See \cite[Section 12.1]{D04} and \cite[Section 10.5]{BDLH05}.)  This leads us to define a new (dual) quantum boundary exponent, 
 \begin{eqnarray}\label{eq:DeltaLj+}
 \Delta^{\kappa}_{L,j}:=\Delta_L^{\kappa}+j\, \mathcal U^{-1}_\kappa(0)=\frac{2L}{\kappa}+j\,\vartheta(\kappa-4) \left(1-\frac{4}{\kappa}\right),
  \end{eqnarray}
 associated with a boundary-attached $L$-star, with a number $j$ of arm-splittings.  %In particular, \emph{putting $j=2$ insertions in between the boundary and the left-most and right-most paths of the $L$-star yields Dirichlet b.c.'s and the ordinary boundary phase transition for non-simple} SLE$_{\kappa>4}$ (Fig \ref{fig:multipletheta}). 

 These dual quantum boundary dimensions now are the seeds to generate Euclidean boundary and bulk scaling exponents via the (dual) KPZ relation \eqref{eq:KPZ},
  \begin{eqnarray}\label{eq:KPZLj}
x^{S,\kappa}_{L,j}:=\mathcal U_\kappa(\Delta^\kappa_{L,j})&=&\frac{1}{4}\Delta^\kappa_{L,j} \left(\kappa \Delta^\kappa_{L,j}+4-\kappa\right)=\frac{\kappa}{4} \Delta^\kappa_{L,j} \Delta^\kappa_{L,j-1},\\ \nonumber &&0\leq j\leq L+1,\\ \nonumber
x^{\kappa}_{L,j}:=2\mathcal V_\kappa(\Delta^\kappa_{L,j})&=&2\,\mathcal U_\kappa\left[\frac{1}{2}\left(\Delta^\kappa_{L,j} +1-\frac{4}{\kappa}\right)\right]=2\,\mathcal U_\kappa\left[\frac{1}{2}\Delta^\kappa_{L,j+1}\right]\\  \label{eq:BtoB}
&=&%\frac{1}{8\kappa}\left[\kappa^2 \Delta_{L,j}^2-(4-\kappa)^2\right]=
\frac{\kappa}{8}\Delta^\kappa_{L,j+1}\Delta^\kappa_{L,j-1},%\\ \nonumber &&
\,\,\,0\leq j\leq L.
\end{eqnarray}
This gives for $\kappa>4$  the set of explicit Euclidean exponents,
%  \begin{eqnarray}\label{eq:xSLj}
%x^{S,\kappa}_{L,j}&=&\frac{\kappa}{4} \Delta_{L,j} \Delta_{L,j-1}=\frac{\kappa}{4}\left[\frac{2L}{\kappa}+j\,\left(1-\frac{4}{\kappa}\right)\right]\left[\frac{2L}{\kappa}+(j-1)\,\left(1-\frac{4}{\kappa}\right)\right],\\ \nonumber
%x^{\kappa}_{L,j}
%&=&\frac{\kappa}{8}\Delta_{L,j+1}\Delta_{L,j-1}\\ \label{eq:xLj}
%&=&\frac{\kappa}{8}\left[\frac{2L}{\kappa}+(j+1)\,\left(1-\frac{4}{\kappa}\right)\right]\times \left[\frac{2L}{\kappa}+(j-1)\,\left(1-\frac{4}{\kappa}\right)\right].
%\end{eqnarray}
  \begin{eqnarray}\nonumber
x^{S,\kappa}_{L,j}&=& \frac{\kappa}{4}\Delta^\kappa_{L,j} \Delta^\kappa_{L,j-1},\,\,\,\,\,\, 0\leq j\leq L+1,\\ \label{eq:xSLj}
&=&\frac{1}{4\kappa}\left[{2L}+j\,\left({\kappa}-{4}\right)\right]\left[{2L}+(j-1)\,\left({\kappa}-{4}\right)\right],\\ \nonumber
&&\\ \nonumber
x^{\kappa}_{L,j}
&=&\frac{\kappa}{8}\Delta^\kappa_{L,j+1}\Delta^\kappa_{L,j-1},\,\,\,\,\,\, 0\leq j\leq L,\\ \label{eq:xLj}
&=&\frac{1}{8\kappa}\left[{2L}+(j+1)\,\left(\kappa-{4}\right)\right] \left[{2L}+(j-1)\,\left(\kappa-{4}\right)\right].
\end{eqnarray}
%\begin{eqnarray}\label{eq:xSLj}
%x^{S,\kappa}_{L,j}&=& \frac{1}{4\kappa}\left[{2L}+j\,\left({\kappa}-{4}\right)\right]\left[{2L}+(j-1)\,\left({\kappa}-{4}\right)\right],\,\,\,\, 0\leq j\leq L+1,\\ \label{eq:xLj}
%x^{\kappa}_{L,j}
%&=&
%\frac{1}{8\kappa}\left[{2L}+(j+1)\,\left(\kappa-{4}\right)\right] \left[{2L}+(j-1)\,\left(\kappa-{4}\right)\right], \,\,\,\, 0\leq j\leq L.
%\end{eqnarray}
These exponents describe for $\kappa>4$ an $L$-star SLE$_\kappa$ with $j$ constraints associated with the conditioning  of non-intersection 
  %or with the mutually bouncing intersection for $\kappa\leq 4$,  
of $j$ pairs of paths or pairs comprised of a path and a boundary line, either in the chordal setting with $0\leq j\leq L+1$, or in the radial or whole-plane setting with $0\leq j\leq L$ \cite{D04,BDLH05}. The $j=0,1$ cases of exponents \eqref{eq:xSLj} appear in \cite{WuZhan17}, and the $j=0,1,2$ cases of exponents \eqref{eq:xLj} in \cite{HWu18}.

For $j=0$, we recover exponents \eqref{eq:KPZL}, \eqref{eq:BtoBL}, which together with the $j=1$ and $j=2$ cases yield the boundary exponents, 
  \begin{eqnarray}\label{eq:xSL0}
  x^{S,\kappa}_{L,j=0}&=& \frac{L}{2\kappa}\left({2L}+4-{\kappa}\right),\\
  \label{eq:xSL1}
  x^{S,\kappa}_{L,j=1}&=& \frac{L}{2\kappa}\left({2L}+{\kappa}-{4}\right),\\
%  x^{\kappa}_{L,j=1}
%&=&
%\frac{1}{8\kappa}\left[{2L}+(j+1)\,\left(\kappa-{4}\right)\right] \left[{2L}+(j-1)\,\left(\kappa-{4}\right)\right],\\
  \label{eq:xSL2}
x^{S,\kappa}_{L,j=2}&=& \frac{1}{2\kappa}\left({L}+{\kappa}-{4}\right)\left({2L}+{\kappa}-{4}\right).
%x^{\kappa}_{L,j=2}
%&=&
%\frac{1}{8\kappa}\left[{2L}+(j+1)\,\left(\kappa-{4}\right)\right] \left[{2L}+(j-1)\,\left(\kappa-{4}\right)\right].
\end{eqnarray}
These exponents will be used later in application to  polymers at the $\Theta$-point in two-dimensions. 
%as the Neumann special transition boundary exponents for $\kappa\leq 4$ and Dirichlet ordinary ones for $\kappa>4$. 
%\textcolor{blue}{This can be simply recapitulated as,
%\begin{eqnarray}
 %x^{S,\kappa \leq 4}_{L}(\mathrm{ord})&=&x^{S,\kappa\leq 4}_{L,j=0},\,\,\,\,
%x^{S,\kappa \leq 4}_{L}(\mathrm{sp})=x^{S,\kappa\leq 4}_{L,j=2};\\ 
%x^{S,\kappa > 4}_{L}(\mathrm{ord})&=&x^{S,\kappa>4}_{L,j=2},\,\,\,\,
%x^{S,\kappa > 4}_{L}(\mathrm{sp})=x^{S,\kappa> 4}_{L,j=0}.
%\end{eqnarray}}
\subsubsection{Duality and path frontiers.} It is known that the \emph{external frontier} of a non-simple SLE$_{\kappa'}$ process $\eta'$ with $\kappa' > 4$ is a form of simple SLE$_\kappa$ process $\eta$ with dual parameter 
$16/\kappa'=:\kappa < 4$ \cite{BD2000,D04,Dubedat2005,Zhan2008}. For instance, one has for their respective Hausdorff dimensions $\mathrm{dim}_H (\eta')=2-x_2^{\kappa'}=1+\kappa'/8$, and $\mathrm{dim}_H (\eta)=1+2/\kappa'=2-x_2^{\kappa}=1+\kappa/8$. Near an ordinary boundary, a similar duality holds. Indeed, for $L=2$, Eqs. \eqref{eq:xSL0} and \eqref{eq:xSL2} yield the identity
\BE\label{eq:dualitysurf}
x^{S,\kappa'}_{L=2,j=2}= \frac{1}{2}\left({\kappa'}-{2}\right)= \frac{1}{\kappa}\left(8-{\kappa}\right)=x^{S,\kappa}_{L=2,j=0}.
\EE  
This means that a non-simple path $\eta'$, anchored at a surface and conditioned not to hit it outside the boundary root, is conformally equivalent to the simple path $\eta$ made by its external frontier   anchored at the same root.  
\subsection{Conformal welding of quantum wedges}\label{sec:confweld}
 \begin{figure}[h!]
\centering
\includegraphics[angle=0,scale =1.52] {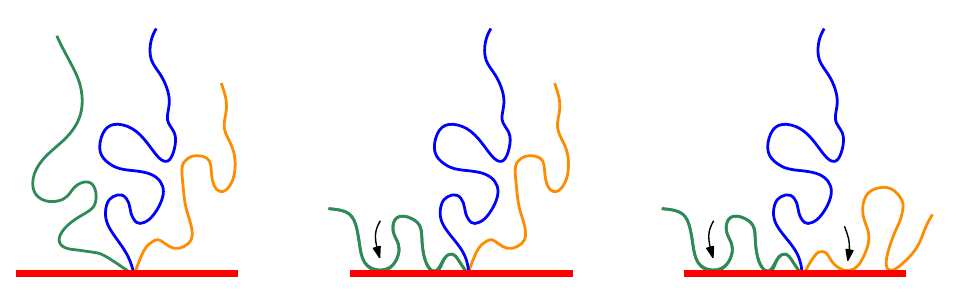}
 \caption{$L=3$ multiple chordal SLE$_{\kappa\leq 4}$ paths in two-dimensions. For $\kappa\leq 4$, the simple paths do not touch each other, nor the surface.  The paths can be conditioned in various ways. {\sl Left:} Ordinary boundary conditions where the paths avoid the surface, with exponent $x_L^{S,\kappa}$. {\sl Middle:} Conditioning one path  (as marked by one arrow) to intersect the surface. {\sl Right:}  Conditioning two paths  (as marked by arrows) to intersect the surface.}
 \label{fig:DtoNbis}
\end{figure}
\subsubsection{Introduction.}In this section, we shall make use of the formalism of conformal welding of quantum wedges and cones in Liouville Quantum Gravity (LQG), as developed  in Refs. \cite{She16} and \cite{DMS2014}.  We shall first use it to condition multiple SLE$_\kappa$  paths for $\kappa\leq 4$, and further show how to recover and interpret the scaling dimensions  \eqref{eq:xSLj} and \eqref{eq:xLj} for $\kappa>4$ in that formalism. 

We shall need the SLE$_\kappa(\rho)$ process  \cite{WW2004a,WW2004b,Dubedat2005}, an important variant of the standard $\mathrm{SLE}_\kappa$ process. In its chordal version in $\mathbb H$, it has a special point marked near the origin on $\partial \mathbb H$ ($0^+$, say), and  roughly speaking, the parameter $\rho$ indicates whether there is an attraction ($\rho <0$) to, or a repulsion ($\rho>0$) from the half-boundary $\mathbb R^+$. When $\rho<\kappa/2-2\leq 0$, the corresponding  curves will touch the boundary; when $\rho > \kappa/2-2$, the curves do not touch the boundary except at the end points, as depicted in Fig. \ref{fig:DtoNbis}. 
This description of the boundary behaviour of cordal SLE$_\kappa(\rho)$ (with a force point located at $0^+$) can be read off from the almost sure Hausdorff dimension of the intersection of its trace $\eta$ with the rightmost part of the boundary.  For $\rho\in(-2\vee \kappa/2-4, \kappa/2-2)$, it is \cite{DubJS09,DubJS10,MillerWu17,WangWu15},
\BE\label{eq:SLErho}
\textrm{dim} (\eta\cap \mathbb R^+)=1-\beta,\,\,\, \beta=\beta (\rho):=\frac{1}{\kappa} (2+\rho)(2+\rho +2 -\kappa/2),  
\EE  
such that $0< \textrm{dim} (\eta\cap \mathbb R^+) <1$ for $\rho\in(-2\vee \kappa/2-4, \kappa/2-2)$. For $\rho > \kappa/2-2$, the trace does not intersect the boundary. 
Similarly, one can define a chordal $\textrm{SLE}_\kappa(\rho_1,\rho_2)$ process with two force points located at $0^-$ and $0^+$.
%Consider then the last case in Fig. \ref{fig:DtoNbis}. Here, we view it as involving a collection of chordal SLE$_\kappa \leq 4$ paths $\eta_i$, $1\leq i\leq L$. The conditional law of each path $\eta_i$, given $\eta_{j}$ for $j\neq i$  is that of an SLE$_\kappa(\rho_i,\rho_{i+1})$ in the component of $\mathbb H \setminus \cup_{j\neq i} \eta_j$ between $\eta_{i-1}$ and $\eta_{i+1}$, with the convention that $\eta_0=\mathbb R^-$ and  $\eta_{L+1}=\mathbb R^+$. Later, we shall mainly be interested in the case $\rho_1\neq 0, \rho_L\neq 0$, with all other $\rho_i=0, i\in \{2,L-1\}$.
\begin{figure}[htbp]
\centering
\includegraphics[angle=0,scale =0.7329] {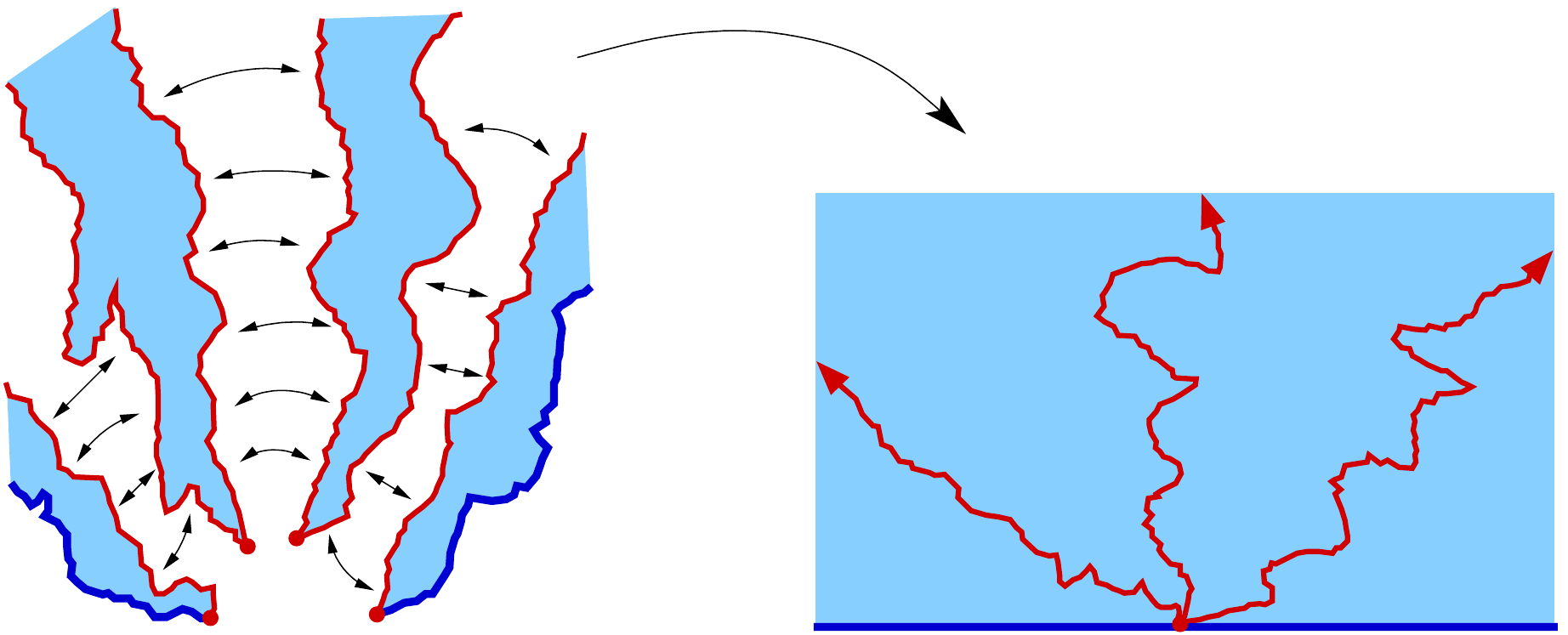}
 \caption{Four quantum wedges conformally welded along boundaries according to their boundary quantum length, and conformally mapped to $\mathbb H$. The images of the interfaces are coupled SLE$_\kappa(\rho_1,\rho_2)$ processes.}
 \label{fig:wedges}
\end{figure}
\subsubsection{Simple SLE paths ($\kappa\leq 4$).}\label{sec:simplepaths}Consider Fig. \ref{fig:DtoNbis}. Here, we view it as depicting a collection of chordal SLE$_{\kappa \leq 4}$ paths $\eta_i$, $1\leq i\leq L$.  For $2\leq i\leq L-1$, the conditional law of each path $\eta_i$, given $\eta_{j}$ for $j\neq i,$  is that of an SLE$_\kappa$ in the component of $\mathbb H \setminus \cup_{j\neq i} \eta_j$ between $\eta_{i-1}$ and $\eta_{i+1}$, whereas the conditional law of  $\eta_{i=1}$ is that  of an SLE$_\kappa(\rho_1)$ in the sector between $\mathbb R^-$ and $\eta_2$, and the conditional law of $\eta_{i=L}$ is that of an SLE$_\kappa(\rho_2)$ in the sector between $\eta_{L-1}$ and $\mathbb R^+$. The marginal law of $\eta_i$ for each $i=1,\cdots,L$ is that of an SLE$_{\kappa}(\rho_1+2(i-1),2(L-i)+\rho_2)$ process. This setting is a slight generalisation of that considered for $\rho_1=\rho_2=0$ in \cite[Appendix B.4, Figure B.2]{DMS2014}. We shall in the sequel  use the terminology and results of Ref. \cite{DMS2014}; for smoother introductions, see \cite{BDICM2014,GHS2019}. 

As shown in Fig. \ref{fig:wedges}, and explained in \cite[Appendix B.4]{DMS2014}, this collection of SLE$_\kappa$ paths can be obtained by first considering in LQG a so-called \emph{quantum wedge} $\mathcal W$, made by gluing together $L+1$ independent quantum wedges  $\mathcal W_i, i=1,\cdots,L+1$, of respective weights $W_1=2+\rho_1$, $W_i=2, i=2,\cdots L$, and $W_{L+1}=2+\rho_2$.The conformal welding of these independent wedges is made according to their $\gamma$-LQG boundary length, followed by a conformal mapping to $\mathbb H$. The resulting $L$ interfaces are then represented by the set $\eta_i, i=1,\cdots,L,$ of SLE$_\kappa$ curves. The LQG parameter $\gamma$ which determines the quantum boundary measure $e^{\gamma h/2} dx$ \cite{DuSh2009,DuSh2011}, where $h$ is an instance of the Gaussian Free Field (GFF) with free boundary conditions, is related to $\kappa$ by the fundamental relation \cite{She16,DuSh2011b,DMS2014},  
\BE\label{eq:gammakappa}
\gamma=\sqrt{\kappa}\wedge \frac{4}{\sqrt{\kappa}}. 
\EE
In term of this Liouville parameter $\gamma$, the standard KPZ relation is \cite{DuSh2009,DuSh2011,BD2010,BDICM2014}
\begin{eqnarray}\label{eq:KPZstandard}
x=U_\gamma(\Delta):=\frac{\gamma^2}{4}\Delta^2+\left(1-\frac{\gamma^2}{4}\right)\Delta,\\ \label{eq:KPZstandardinv}
\Delta=U^{-1}_\gamma(x)=\frac{1}{\gamma}\left[\sqrt{4x+a^2_\gamma} -a_\gamma\right],\,\,\,a_\gamma:=\frac{2}{\gamma}-\frac{\gamma}{2},
\end{eqnarray}
and is identical to Eq. \eqref{eq:KPZ} in the case $\kappa \leq 4$, where $\gamma^2=\kappa$.  This is in particular the case we are dealing with in this section. 

The weight of the wedge $\mathcal W$ resulting from the welding of independent wedges is simply the \emph{sum} of their individual weights, 
\BE\label{eq:WWi}
W=\sum_{i=1}^{L+1}W_i=2(L+1)+\rho_1+\rho_2.
\EE 
According to the LQG formalism (see \cite[Section 1.1 and Appendix B]{DMS2014}),  such a wedge $\mathcal W$ of weight $W$ is an $\alpha$-\emph{quantum wedge}  where
\BE\label{eq:Walpha}\alpha= \gamma+2/\gamma-W/\gamma,
\EE
 such that the Gaussian free field $h$ is locally $\alpha$-\emph{thick} (corresponding to a singularity $-\alpha \log |\cdot |$ at the origin). As explained in \cite[Eq. (63)]{DuSh2011} and in \cite{DuSh2011b}, a ``quantum typical'' point in a random fractal of quantum scaling exponent $\Delta$ is then a point near which the field is $\alpha$-thick, with  
\BE\label{eq:alphaDelta}
\alpha=\gamma(1-\Delta).
\EE 
This in turn yields the simple wedge-boundary scaling exponent relation \cite{DMS2014}, 
\BE\label{eq:WDelta}
W=2+\gamma^2\Delta.
\EE
Here the quantum typical point is the origin of the $L$-multiple SLE curves, the marked point of quantum wedge $\mathcal W$ of weight $W$ \eqref{eq:WWi}. It has for quantum boundary scaling exponent $\Delta=\Delta_L^\kappa(\rho_1,\rho_2)$, where  
\BE\label{eq:DeltaL}
\Delta_L^\kappa(\rho_1,\rho_2):=\frac{1}{\gamma^2} (W-2)=\frac{2L}{\kappa}+\frac{1}{\kappa}(\rho_1+\rho_2),
\EE
where use was made of $\gamma^2=\kappa$ from Eq. \eqref{eq:gammakappa} for $\kappa \leq 4$. 
%$\gamma+2/\gamma-W/\gamma=\gamma+2/\gamma-[2(L+1)+\rho_1+\rho_2]/\gamma$.
By using the KPZ relation \eqref{eq:KPZ}, we therefore get the set of Euclidean boundary conformal weights,
\BE\label{eq:xLkrr}
x_L^{S,\kappa}(\rho_1,\rho_2):=\mathcal U_{\kappa}(\Delta_L^\kappa(\rho_1,\rho_2))=\frac{1}{4\kappa}(2L+\rho_1+\rho_2)(2L+\rho_1+\rho_2+4-\kappa),
\EE 
which correspond to the boundary scaling behaviour of the set of $L$-multiple chordal SLE$_\kappa$ paths $\eta_i, i=1,\cdots,L$ (Figs. \ref{fig:DtoNbis} and \ref{fig:wedges}). As already shown in \cite[Appendix B.6]{DMS2014}, we recover  for $L=2, \rho_1=\rho_2=\rho$, the scaling exponent $x_2^{S,\kappa}(\rho,\rho)=\beta(\rho)$ that gives the Hausdorff dimension \eqref{eq:SLErho} of the boundary intersection of SLE$_\kappa(\rho)$. 
\begin{figure}[htbp]
\centering
\includegraphics[angle=0,scale =1.329] {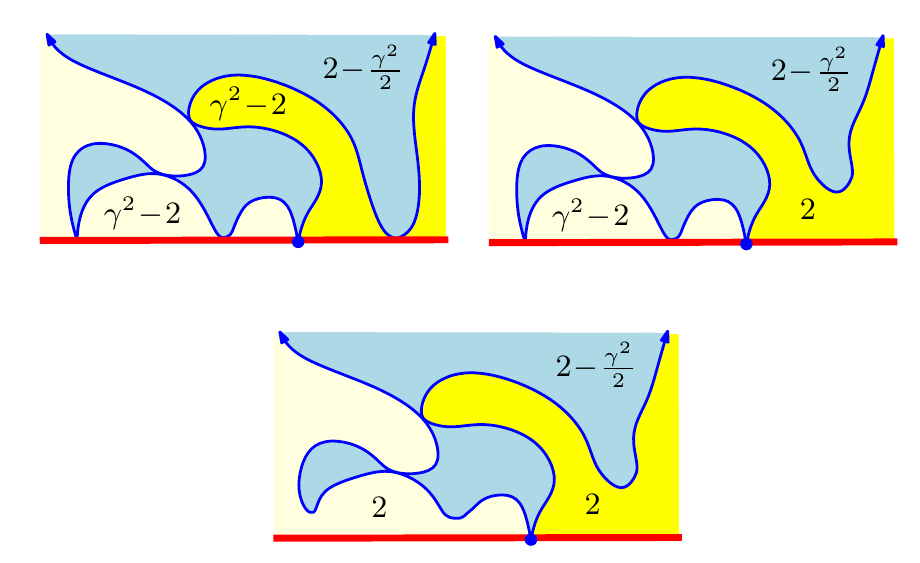}
 \caption{Boundary cases for $L=1$ and $j\in \{0,1,2\}$ in Section \ref{sec:nonsimplepaths}: A quantum wedge $\mathcal W_1$ of weight $W_1=2-\gamma^2/2$ (in blue) is associated with the region surrounded by a non-simple SLE$_{\kappa>4}$ path  $\eta_1$, whose left and right boundaries are represented as thick lines with arrows (in dark blue). Wedges ${\widetilde{\mathcal W}}_{\ell=1,2}$ (in light or bright yellow) have weights $\widetilde W_{\ell=1,2}$ equal to either $\gamma^2-2$ or $2$ depending on whether $\eta_1$ is unconditioned or conditioned not to intersect the corresponding half-boundary, respectively.}
 \label{fig:boundarywedges}
\end{figure}
\subsubsection{Non-simple SLE paths near a boundary ($\kappa>4$).}\label{sec:nonsimplepaths} Consider again Fig. \ref{fig:multiplethetabis}, and the collection it depicts of SLE$_{\kappa >4}$ non-simple paths $\eta_i,i=1,\cdots,L$, in clockwise order, and with the further convention that $\eta_0=\mathbb R^-$ and $\eta_{L+1}=\mathbb R^+$.  Here, following \cite[Appendices B.2, B.4, B.5]{DMS2014} we view their configuration in $\mathbb H$ as resulting from the conformal welding of a set of $2L+1$ independent quantum wedges, followed by a conformal map to $\mathbb H$  (Fig. \ref{fig:boundarywedges}). These wedges are of two main types. Wedges $\mathcal W_i, i=1,\cdots,L,$  represent the regions surrounded by the non-simple paths $\eta_i, i=1,\cdots,L$, i.e., the wedges located in between the left and right boundary of each path $\eta_i$. Each of them has weight $W_i=2-\gamma^2/2$ \cite{DMS2014}. They are separated from the boundary and from each other by another set of wedges $\widetilde{\mathcal W}_\ell, \ell=1,\cdots L+1$, interspersed in between wedges $\mathcal W_i, i=1,\cdots,L$. The wedge $\widetilde{\mathcal W}_{\ell=i}$ represents the quantum space in between the right boundary of $\eta_i$ and the left boundary of $\eta_{i+1}$. If the latter two are \emph{free} to be in contact, the weight of $\widetilde{\mathcal W}_{i}$ is ${\widetilde W}_{i}=\gamma^2-2$, whereas if  $\eta_i$ and $\eta_{i+1}$ are \emph{conditioned} not to hit each other, the weight is ${\widetilde W}_{i}=2$ \cite[Appendix 4]{DMS2014}. Equivalently, the  conditional law of $\eta_i$ is that of an SLE$_\kappa (\kappa-4)$ if it is simply conditioned not to intersect $\eta_{i-1}$, and that of an SLE$_\kappa(\kappa-4,\kappa-4)$ if it does not intersect both $\eta_{i-1}$ and $\eta_{i+1}$ \cite{Dubedat2005,MS2016}.  Let us now assume that there is a number $j$, $0\leq j\leq L+1,$ of such conditionings. In Fig. \ref{fig:multiplethetabis}, these conditionings are represented by arrows located in between SLE paths, while in Fig. \ref{fig:boundarywedges}, they correspond to the presence of wedges of weight $2$. 

The resulting wedge $\mathcal W$ has for its weight the sum of individual weights, 
\begin{eqnarray}\label{eq:Wdense}
W&=&L(2-\gamma^2/2)+ (L+1-j)(\gamma^2-2)+2j\\ \nonumber
&=&L\gamma^2/2+(4-\gamma^2)j+\gamma^2-2.
\end{eqnarray}
The relation \eqref{eq:WDelta} between a quantum wedge weight  $W$ and the corresponding boundary quantum scaling exponent $\Delta$ yields  for the weight \eqref{eq:Wdense} a boundary quantum scaling exponent $\Delta=\Delta_{L,j}$, 
\begin{eqnarray}\label{eq:tildeDeltaLj}
\Delta_{L,j}&:=&\frac{1}{\gamma^2}(W-2)=\frac{L}{2}+(j-1)\left(\frac{4}{\gamma^2}-1\right).
\end{eqnarray}
Recall that exponents $\Delta^{\kappa}_{L,j}$ in Eq. \eqref{eq:DeltaLj+} are for $\kappa>4$ \emph{dual} quantum dimensions. By definition \cite{D04,BDLH05,BD2010}, the dual $\widetilde \Delta$ of a quantum scaling exponent $\Delta$ obeys 
\BE\label{eq:dualdim}
\gamma(1-\Delta)=(4/\gamma)(1-\widetilde \Delta).
\EE 
The relation \eqref{eq:WDelta} between a quantum wedge weight  $W$ and the corresponding boundary quantum scaling exponent $\Delta$ thus becomes for its dual $\widetilde \Delta$ \cite[Appendix B.3]{BDICM2014}, 
\BE\label{eq:WDeltadual}
W=\gamma^2-2+4\widetilde \Delta.
\EE 
Therefore,  the weight \eqref{eq:Wdense} yields a dual boundary quantum scaling exponent  $\widetilde \Delta=\widetilde \Delta_{L,j}$, where 
\begin{eqnarray}\label{eq:ttildeDeltaLj}
%\widetilde\Delta^{\kappa}_{L,j}&:=&\frac{1}{\gamma^2}(W-2)=L/2+(j-1)(4/\gamma^2-1),\\ \nonumber
%&=&L/2+(j-1)(\kappa/4-1)\\
\widetilde\Delta_{L,j}&:=&\frac{1}{4}(W+2-\gamma^2)=\frac{\gamma^2}{8}L+j \left(1-\frac{\gamma^2}{4}\right) .
\end{eqnarray}
Using now \eqref{eq:gammakappa} for $\kappa>4$ gives  $\gamma^2=16/\kappa$, and we get the pair of dual boundary quantum scaling dimensions,
\begin{eqnarray}\label{eq:standardDeltaLj}
\Delta_{L,j}&=&\frac{L}{2}+(j-1)\left(\frac{\kappa}{4}-1\right),\\\label{eq:dualDeltaLj}
\widetilde\Delta_{L,j}&=&\frac{2L}{\kappa}+j\left(1-\frac{4}{\kappa}\right)=\Delta^{\kappa}_{L,j},
\end{eqnarray}
where, as expected, Eq. \eqref{eq:dualDeltaLj} coincides with Eq. \eqref{eq:DeltaLj+} for the dual quantum dimension.

Finally, using either the standard KPZ relation \eqref{eq:KPZstandard} for the standard quantum dimension \eqref{eq:standardDeltaLj}, or equivalently, its dual version \eqref{eq:KPZ} for the dual dimension \eqref{eq:dualDeltaLj}, we get 
\BE\label{eq:xLjk}
x^{S,\kappa}_{L,j}=U_\gamma\left(\Delta_{L,j}\right)=\mathcal U_\kappa\left(\Delta^{\kappa}_{L,j}\right)=\frac{1}{4\kappa}\left[{2L}+j\,\left({\kappa}-{4}\right)\right]\left[{2L}+(j-1)\,\left({\kappa}-{4}\right)\right],
\EE
which coincides with Eq. \eqref{eq:xSLj}.
\begin{figure}[htbp]
\centering
\includegraphics[angle=0,scale =1.329] {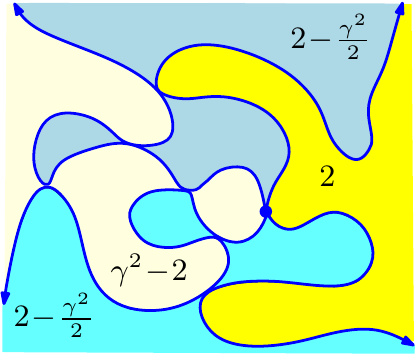}
 \caption{Bulk case for $L=2$ and $j=1$ in Section \ref{sec:bulknonsimplepaths}: Two quantum wedges $\mathcal W_{i=1,2}$ of weight $W_{i=1,2}=2-\gamma^2/2$ (in blue and cyan) are associated with the regions surrounded by two non-simple SLE$_{\kappa>4}$ paths  $\eta_{i=1,2}$, whose left and right boundaries are represented as thick lines with arrows (in dark blue). Wedge  ${\widetilde{\mathcal W}}_{1}$ (in light yellow) has weight $\widetilde W_{1}=\gamma^2-2$ because the left boundary of $\eta_1$ and the right boundary of $\eta_2$ are allowed to intersect, while wedge  ${\widetilde{\mathcal W}}_{2}$ (in bright yellow) has weight $\widetilde W_{2}=2$ because the right boundary of $\eta_1$ and the left boundary of $\eta_2$ are conditioned not to intersect.}
 \label{fig:bulkwedges}
\end{figure}

\subsubsection{Non-simple SLE paths  in the bulk ($\kappa > 4$).}\label{sec:bulknonsimplepaths} Consider the local behaviour of a collection of $L$ successive  whole-plane SLE$_{\kappa>4}$ processes $\eta_i$, $i=1,\cdots, L,$ near a quantum typical multiple point $z_0$ of order $L$, with a number $j$, $0\leq j\leq L$, of conditionings of pairs of successive, non-intersecting paths. It can be represented by gluing together $2L$ independent wedges of alternating weights $2-\gamma^2/2$, for the region surrounded by each $\eta_i$, and either $2$ or $\gamma^2-2$ for the region between $\eta_i$ and $\eta_{i+1}$, depending on whether the pair $(\eta_i,\eta_{i+1})$ is conditioned not to intersect or is not conditioned (Fig \ref{fig:bulkwedges}; see also \cite[Appendix B.5]{DMS2014} in the $j=0$ case). This yields a total weight,
\BE\label{eq:coneweight}
W=L\left(2-\frac{\gamma^2}{2}\right)+2j+\left(L-j\right) \left(\gamma^2-2\right)=L\frac{\gamma^2}{2}+j(4-\gamma^2).
\EE 
By the conformal welding theory in Liouville quantum gravity \cite[Proposition (7.16)]{DMS2014}, gluing together these $2L$ wedges yields a so-called  \emph{$\alpha$-quantum cone} associated with a local singularity $-\alpha \log |\cdot-z_0|$ near the $L$-multiple point $z_0$, where 
\BE\nonumber
\alpha:=Q-\frac{W}{2\gamma},\,\,\, Q:=\frac{\gamma}{2}+\frac{2}{\gamma}.
\EE
Its explicit value is here,
\BE\label{eq:alphacone}
\alpha=\frac{2}{\gamma}+\frac{\gamma}{2}-\frac{\gamma}{4}L-\left(\frac{2}{\gamma}-\frac{\gamma}{2}\right) j.
\EE
Finally, let $\Delta=\Delta^{\circ}_{L,j}$ be the bulk quantum dimension associated with this $\alpha$-cone so that $\alpha=\gamma(1-\Delta)$. We find
\BE\label{eq:DeltacircLj}
\Delta^{\circ}_{L,j}=\frac{L}{4}+\frac{1}{2}(j-1)\left(\frac{4}{\gamma^2}-1\right).
\EE
Because of Definition \eqref{eq:dualdim}, the dual dimension $\widetilde \Delta$ of $\Delta$ obeys $\alpha=(4/\gamma)(1-\widetilde \Delta)$, so that the dual of dimension \eqref{eq:DeltacircLj} reads  with the help of \eqref{eq:alphacone}, 
\BE\label{eq:DeltacircLjdual}
\widetilde \Delta^{\circ}_{L,j}=1-\frac{\gamma}{4}\alpha=\left(\frac{\gamma}{4}\right)^2L+ \frac{1}{2} (j+1) \left(1-\frac{\gamma^2}{4}\right).
\EE
These results hold for $\kappa>4$ and $\gamma^2=16/\kappa$. We thus arrive at the pair of dual bulk quantum dimensions,
\begin{eqnarray}\label{eq:standardDeltaLjcirc}
\Delta^{\circ}_{L,j}&=&\frac{L}{4}+\frac{1}{2}(j-1)\left(\frac{\kappa}{4}-1\right),\\
\label{eq:dualDeltaLjcirc}
\widetilde\Delta^{\circ}_{L,j}&=&\frac{L}{\kappa}+\frac{1}{2}(j+1)\left(1-\frac{4}{\kappa}\right)=:\Delta^{\circ\,\kappa}_{L,j},
\end{eqnarray}
where we used the dual version to define $\Delta^{\circ\,\kappa}_{L,j}$ in \eqref{eq:dualDeltaLjcirc}, in accordance with the definition of the dual dimension $\Delta^{\kappa}_{L,j}$ in Eq. \eqref{eq:dualDeltaLj}.  We thus have the relationship between the quantum  boundary and bulk dimensions \eqref{eq:standardDeltaLj} and \eqref{eq:standardDeltaLjcirc}, or between their respective duals  \eqref{eq:dualDeltaLj} and \eqref{eq:dualDeltaLjcirc},
\begin{eqnarray}
&\Delta_{L,j}&=2\Delta^{\circ}_{L,j},\\
\Delta^{\kappa}_{L,j}=&\widetilde \Delta_{L,j}&=2\widetilde \Delta^{\circ}_{L,j}-(1-\gamma^2/4)=2 \Delta^{\circ\,\kappa}_{L,j}-(1-4/\kappa),
\end{eqnarray} 
in full agreement with previous results for the dense phase of the $O(n)$ model and for SLE$_{\kappa>4}$ non-simple paths. (See  \cite[Section 10.3, Equations (10.23) and (10.24); Section 11.1, Equation (11.2)]{D04}.)

Finally, using either the standard KPZ relation \eqref{eq:KPZstandard} for the standard bulk quantum dimension  \eqref{eq:DeltacircLj} or \eqref{eq:standardDeltaLjcirc}, or equivalently, its dual versions \eqref{eq:KPZ} and \eqref{eq:BtoBfirst} for the dual dimensions \eqref{eq:dualDeltaLjcirc} and \eqref{eq:dualDeltaLj}, we get the corresponding set of bulk Euclidean scaling dimensions,
\begin{eqnarray}\nonumber
x^{\kappa}_{L,j}&=&2U_{\gamma}\left(\Delta^{\circ}_{L,j}\right)=2\,\mathcal U_\kappa\left(\Delta^{\circ\,\kappa}_{L,j}\right)=\mathcal V_\kappa\left(\Delta^{\kappa}_{L,j}\right) \\ \label{eq:xLjkbulk}
&=&\frac{1}{8\kappa}\left[{2L}+(j+1)\,\left({\kappa}-{4}\right)\right]\left[{2L}+(j-1)\,\left({\kappa}-{4}\right)\right],
\end{eqnarray}
which coincides with Eq. \eqref{eq:xLj}, as expected.

\subsection{Special boundary conditions} In this section, we shall recast results known for the special surface transition of the $O(n)$ model at its dilute critical point, in terms of the formalism of multiple SLE$_\kappa(\rho_1,\rho_2)$ paths, as  described in Section \ref{sec:simplepaths} above. 
For the special surface transition the boundary $L$-star exponents $x_L^{S}(\textrm{sp})$ are given by \cite{FS94,BY95,BY95bis,YungBat95}
\BE\label{eq:xLspe}
x_L^{S}(\textrm{sp})=\frac{1}{4}g(L+1)^2-\frac{3}{2}(L+1)+\frac{1}{4g}[9-(g-1)^2],
\EE
with $g$ the Coulomb gas coupling constant such that $n=-2\cos \pi g$, with $1\leq g\leq 2$ at the dilute critical point. They are also the conformal weights $h_{L+1,3}$ in terms of the Kac table $h_{p,q}:= \frac{1}{4g}[(gp-q)^2-(g-1)^2].$ 
For Schramm-Loewner SLE$_\kappa$ paths, we have here $\kappa=4/g$ with $2\leq \kappa\leq 4$, and get
\BE
\label{eq:xLspk}
x_L^{S}(\textrm{sp})=\frac{1}{4\kappa}(2L-\kappa)(2L+4-2\kappa)=x_L^{S,\kappa}\left(\rho_1=-\frac{\kappa}{2},\rho_2=-\frac{\kappa}{2}\right),
%x_L^{S,\kappa}\left(\rho_1=-{\kappa}/{2},\rho_2=-{\kappa}/{2}\right),
\EE
in terms of the multiple SLE$_\kappa(\rho_1,\rho_2)$ exponents \eqref{eq:xLkrr}. We remark that according to \eqref{eq:SLErho} the left- and rightmost paths of the $L$-star start to touch the boundary for values $\rho_1=\rho_2=\kappa/2-2$, whereas the special transition occurs for $\rho_1=\rho_2=-\kappa/2$, which for the range $2\leq \kappa\leq 4$ corresponding to the dilute $O(n)$ model, is a \emph{lower} value of the $\rho$ parameter. In terms of that measure, the attraction towards the surface at the special transition is therefore stronger than for a simple contact of the paths. The value $\rho=-\kappa/2$ agrees with that found in the study of the so-called anisotropic special transition \cite{Dieh86,DE82,DE84}, for the onset value $r=1$ of the anisotropy parameter \cite{DubJS09,DubJS10,Duba2010}. 

The Euclidean special boundary exponents \eqref{eq:xLspk} are the images by the KPZ function  \eqref{eq:KPZ} with $\gamma=\sqrt{\kappa}$ or \eqref{eq:KPZstandard},
\BE\label{eq:KPZsp}
x_L^{S}(\textrm{sp})=U_\gamma\big(\Delta^S_L(\textrm{sp})\big)=\mathcal U_\kappa\big(\Delta^S_L(\textrm{sp})\big),
\EE
 of the special quantum boundary dimensions, 
\BE\label{eq:DeltaLsp}
\Delta^S_L(\textrm{sp}):=\frac{2L}{\kappa}-1,
\EE
which thus appear as obtained by a simple $-1$ shift from the ordinary quantum boundary dimensions \eqref{eq:deltaL} $\Delta^S_L(\textrm{or}):=\Delta_L^{\kappa}=2L/\kappa$. (For a related   treatment by random matrix techniques of boundary conditions in the $O(n)$ model on a random lattice, see \cite{Bourgine2010}.)

\subsection{Mixed boundary conditions}
Mixed boundary conditions for the dilute critical $O(n)$ model have been considered in \cite{BY95bis,YungBat95}, where ordinary and special boundary conditions act on either side of the origin on the real axis.  The local scaling property of an $L$-arm star anchored at the origin are then associated with Euclidean mixed boundary exponents $x_L^{S}(\textrm{or}\bullet\textrm{sp})$ that are simply given by the ordinary ones, $x_L^{S}(\textrm{or})$, shifted by $-L/2$,
\begin{eqnarray}\label{eq:or}
&&x_L^{S}(\textrm{or})=\frac{1}{4}gL^2+\frac{1}{2}(g-1)L, \\
\label{eq:orsp}
&&x_L^{S}(\textrm{or}\bullet\textrm{sp})=x_L^{S}(\textrm{or})-\frac{L}{2}=\frac{1}{4}gL^2+\frac{1}{2}(g-2)L. 
\end{eqnarray}
Here again, $1\leq g\leq 2$, and these exponents are for SLE$_\kappa$ paths, with $2\leq\kappa=4/g\leq 4$,
\begin{eqnarray}\label{eq:ork}
x_L^{S}(\textrm{or})=x^{S,\kappa}_L=\frac{L}{2\kappa} \left(2L+4-\kappa\right),\\
\label{eq:orspk}
x_L^{S}(\textrm{or}\bullet\textrm{sp})=x_L^{S}(\textrm{or})-\frac{L}{2}=\frac{L}{\kappa} \left(L+2-\kappa\right). 
\end{eqnarray}
Interestingly, there is no natural, nor simple way to write the mixed exponents  \eqref{eq:orspk} in terms of the multiple SLE$_\kappa(\rho_1,\rho_2)$ exponents \eqref{eq:xLkrr}, except, as we shall see below, for the SAW value $\kappa=8/3$ of the SLE parameter. Nevertheless, let us introduce a \emph{modified} KPZ function $\widehat U_\gamma$, to be compared with $U_\gamma$ \eqref{eq:KPZstandard},
\begin{eqnarray}\label{eq:KPZmod}
x=\widehat U_\gamma(\Delta):=\frac{\gamma^2}{4}\Delta^2+\left(1-\frac{\gamma^2}{2}\right)\Delta,\\ \label{eq:KPZmodinv}
\Delta=\widehat U^{-1}_\gamma(x)=\frac{1}{\gamma}\left[\sqrt{4x+\hat a^2_\gamma} -\hat a_\gamma\right],\,\,\,\hat a_\gamma:=a_\gamma-\frac{\gamma}{2}=\frac{2}{\gamma}-{\gamma}.
\end{eqnarray}
%\BE\label{eq:KPZmod}
%\widehat U_\gamma(\Delta):=\frac{\gamma^2}{4}\Delta^2+\left(1-\frac{\gamma^2}{2}\right)\Delta,
%\EE
We then observe that the exponents \eqref{eq:orspk} are the images by $\widehat U_\gamma$ of the quantum boundary dimensions \eqref{eq:deltaL} $\Delta=\Delta_L^{S,\kappa}=2L/\kappa$, with, as before, $\gamma^2=\kappa$. We thus have in parallel,
\begin{eqnarray}\label{eq:orkU}
x_L^{S}(\textrm{or})=U_\gamma\big(2L/\kappa\big),\,\,\,\,%\\ %=\frac{L}{2\kappa} \left(2L+4-\kappa\right),\\
%\label{eq:orspkU}
x_L^{S}(\textrm{or}\bullet\textrm{sp})=\widehat U_\gamma\big(2L/\kappa\big).
%=x_L^{S}(\textrm{or})-\frac{L}{2}=\frac{L}{\kappa} \left(L+2-\kappa\right). 
\end{eqnarray}
In the LQG rigorous approach to the KPZ relation developed in \cite{DuSh2009,DuSh2011,BD2010,BDICM2014}, the boundary Liouville quantum measure, formally written as     
$\mu_h=e^{\gamma h/2}dx$ in terms of the free boundary GFF $h$,   
is properly defined as the weak limit \cite{DuSh2011}, 
\BE\label{eq:muh}
\mu_h(x) :=\lim_{\varepsilon\to 0} \mu_{h_\varepsilon}(x):= \lim_{\varepsilon\to 0} \varepsilon^{\gamma/4} e^{\gamma h_\varepsilon(x)/2}dx,
\EE
where $h_\varepsilon(x)$ is the average value of $h$ on a semicircle of radius $\varepsilon$ centered at $x\in \partial \mathbb H$. The renormalisation factor in \eqref{eq:muh} actually is 
$\varepsilon^{\gamma/4}=1/\left(\mathbb E \left[e^{(\gamma/2) h_\varepsilon(x)}\right]\right)$, since on the boundary, $\textrm{Var}\, h_\varepsilon(x)=-2\log \varepsilon$. By following the proof of boundary KPZ  in  \cite[Section 6]{DuSh2011} or \cite[Section 18.5]{BD2010} \cite{BDICM2014}, one can then show that the modified KPZ function \eqref{eq:KPZmod}  with the shift $a_\gamma\mapsto \hat a_\gamma=a_\gamma-\gamma/2$ in \eqref{eq:KPZmodinv}, is precisely  obtained by a deterministic shift of the GFF $h$ in the Liouville quantum boundary measure,  as $h\mapsto \hat h:=h-(\gamma/2) \log|\cdot |$. In turn, this shift exactly suppresses in $\mu_{\hat h_\varepsilon} (0)$ the renormalisation factor, making the limit boundary measure highly singular at the origin, which is precisely where the boundary conditions change.

%\textcolor{red}{The fact that a modified KPZ relation occurs in the case of mixed b.c.'s is also vindicated from the conformal field theory point of view advocated in \cite{YungBat95}. The standard (here inverse) KPZ relation \eqref{eq:KPZstandardinv} can be written in terms of the central charge  of the critical $O(n)$ model, $c=1-6(1-g)^2/g$ with $g=4/\gamma^2$, as,
%$$ \Delta=U^{-1}_\gamma(x)=\frac{1}{\gamma}) \left[\sqrt{4x+\frac{1-c}{6}}-\sqrt{\frac{1-c}{6}}\right]$$
%$$ \Delta=U^{-1}_\gamma(x)=({1}/{\gamma}) \left[\sqrt{4x+(1-c)/6}-\sqrt{(1-c)/6}\right].$$
%In Ref. \cite{YungBat95} an \emph{effective} central charge $\hat c=c-24\, h_{1,2}=1-6(2-g)^2/g$ is obtained in the case of mixed b.c.'s, and we observe that the modified (inverse) KPZ function \eqref{eq:KPZmodinv} is given by precisely the same formula as above, but now with that effective central charge $\hat c$ instead of $c$,
%$$ \Delta=\widehat U^{-1}_\gamma(x)=({1}/{\gamma}) \left[\sqrt{4x+(1-\hat c)/6}-\sqrt{(1-\hat c)/6}\right].$$}

 \subsection{Self-avoiding walks}
Let us now turn to the case of polymers, which can be either SAWs, or Brownian paths, or $\Theta$-point polymer chains.
\subsubsection{$L$-walk exponents.} The case of SAWs, or the physical case of polymer chains in a good solvent, is well-known to correspond to the dilute critical point of the $O(n=0)$ model in two-dimensions \cite{PGG72,PGG79,JdC89,Nienhuis82,Nienhuis87}, with its scaling limit conjectured \cite{D04,LSW2004} to be SLE$_\kappa$ for $\kappa=8/3$, in agreement with Eqs. \eqref{eq:xnorm}, \eqref{eq:KPZL} and \eqref{eq:BtoBL}.  Let us first return to mixed boundary exponents. Because of expression \eqref{eq:xLspk} for the special ones, and from the comparison of the last two cases in Fig. \ref{fig:DtoNbis}, 
it would at first seem natural to expect the mixed boundary exponents to be given by \eqref{eq:xLkrr} for $\rho_1=-\kappa/2$ and $\rho_2=0$, as  
\BE\label{eq:xLkr0}
x_L^{S,\kappa}(-\kappa/2,0)=\frac{1}{4\kappa}(2L-\kappa/2)(2L+4-3\kappa/2),
\EE 
but the latter exponents clearly differ from \eqref{eq:orspk}. However, for the particular case of SLE$_{\kappa=8/3}$, that in particular enjoys the restriction property \cite{WW2004a}, that identification holds, and we 
do have for \eqref{eq:orspk} at $\kappa=8/3$, $x^{S}_L(\mathrm{o\bullet s})=x^{S,\kappa=8/3}_{L}\left(-{4}/{3},0\right)$. For SAWs, we thus have the explicit set of bulk and boundary exponents, 
%\begin{eqnarray}\label{eq:xSLj8/3}
%x^{S,\kappa=8/3}_{L,j}&=& \frac{1}{6}\left(\frac{3}{2}L-j\right)\left(\frac{3}{2}L-(j-1)\right),\,\,\,\, 0\leq j\leq L+1,\\ \label{eq:xLj}
%x^{\kappa=8/3}_{L,j}
%&=&
%\frac{1}{12}\left(\frac{3}{2}L-(j+1)\right)\left(\frac{3}{2}L-(j-1)\right), \,\,\,\, 0\leq j\leq L.
%\end{eqnarray}
%\textcolor{red}{ \begin{eqnarray}\label{eq:xSLj8/3bis}
%x^{S,\kappa=8/3}_{L,j}&=& \frac{1}{24}\left({3}L-2j\right)\left({3}L-2(j-1)\right),\,\,\,\, 0\leq j\leq L+1,\\ \label{eq:xLj}
%x^{\kappa=8/3}_{L,j}
%&=&
%\frac{1}{48}\left({3}L-2(j+1)\right)\left({3}L-2(j-1)\right), \,\,\,\, 0\leq j\leq L.
%\end{eqnarray}
%For $j=0$, we recover the standard multiple SAW exponents \eqref{eq:xnorm}, and together with $j=1,2$ we get 
%\begin{eqnarray}\label{eq:xSLj8/3ter}
%x^{S}_L(\mathrm{or})=x^{S,\kappa=8/3}_{L}= \frac{L}{8}\left({3}L+2\right),\\
%x^{S}_L(\mathrm{o\bullet s})=x^{S,\kappa=8/3}_{L}\left(\rho_1=-\frac{\kappa}{2},\rho_2=0\right)= \frac{L}{8}\left({3}L-2\right),\\
%x^{S}_L(\mathrm{sp})=x^{S,\kappa=8/3}_{L}\left(\rho_1=-\frac{\kappa}{2},\rho_2=-\frac{\kappa}{2}\right)= \frac{1}{24}\left({3}L-4\right)\left({3}L-2\right).
%\end{eqnarray}
%\begin{eqnarray}\label{eq:xSLj8/3ter}
%x^{S}_L(\mathrm{or})=x^{S,\kappa=8/3}_{L}= \frac{L}{8}\left({3}L+2\right),\\
%x^{S}_L(\mathrm{o\bullet s})=x^{S,\kappa=8/3}_{L}\left(\rho_1=-{\kappa}/{2},\rho_2=0\right)= \frac{L}{8}\left({3}L-2\right),\\
%x^{S}_L(\mathrm{sp})=x^{S,\kappa=8/3}_{L}\left(\rho_1=-{\kappa}/{2},\rho_2=-{\kappa}/{2}\right)= \frac{1}{24}\left({3}L-4\right)\left({3}L-2\right).
%\end{eqnarray}
\begin{eqnarray}\label{eq:xL8/3}
x_L=x^{\kappa=8/3}_{L}=\frac{1}{48}(3L-2)(3L+2),\\
\label{eq:xSLj8/3ter}
x^{S}_L(\mathrm{or})=x^{S,\kappa=8/3}_{L}=x^{S,\kappa=8/3}_{L}\left(0,0\right)= \frac{L}{8}\left({3}L+2\right),\\ \label{eq:xLos8/3}
x^{S}_L(\mathrm{o\bullet s})=x^{S,\kappa=8/3}_{L}\left(-{\kappa}/{2},0\right)= \frac{L}{8}\left({3}L-2\right),\\ \label{eq:xLss8/3}
x^{S}_L(\mathrm{sp})=x^{S,\kappa=8/3}_{L}\left(-{\kappa}/{2},-{\kappa}/{2}\right)= \frac{1}{24}\left({3}L-4\right)\left({3}L-2\right).
\end{eqnarray}
As indicated before, these exponents are, respectively, the SAW bulk exponents \cite{Nienhuis82, Nienhuis87,S86,S87,BD1999,D04}, the \emph{ordinary} surface exponents of Refs. \cite{DS86,BS93,BD1999,D04}, the \emph{mixed ordinary-special} exponents of Refs. \cite{BY95bis,YungBat95}, and the SAW \emph{special} surface exponents \eqref{eq:xsurfsp} of Refs. \cite{FS94,BY95,BY95bis,YungBat95}. 
\subsubsection{Crossover exponent at the special transition.} The crossover exponent $\phi$ is by definition such that the number of adsorbed vertices $N_{\textrm{ads}}$ of a SAW, or of a self-avoiding polygon (SAP), of length $N$ scales at the special transition as  $N_{\textrm{ads}}\propto N^\phi$. Since the early works on the special transition of the two-dimensional critical $O(n)$ model \cite{BEG89,BE94,FS94}, it has been known that  $\phi=1/2$ for any $n$. Consider then a partly adsorbed SAW (or SAP) $\eta$, of Euclidean size $R$. We have the scaling laws $N_{\textrm{ads}}\propto R^{D_{\textrm{ads}}}$ and $N\propto R^{D}$,  where the special surface Hausdorff dimension $D_{\textrm{ads}}:=\textrm{dim}(\eta\cap \mathbb R)$ and bulk Hausdorff dimension $D:=\textrm{dim}(\eta)$ are respectively given by  Eq. \eqref{eq:xLss8/3} as $D_{\textrm{ads}}=1-x^{S}_2(\mathrm{sp})=2/3$, and by Eq. \eqref{eq:xL8/3} as $D=2-x^{\kappa=8/3}_2=4/3$. These results are in complete agreement with the expected crossover relation $D_{\textrm{ads}}=\phi D=D/2$.
 \subsection{Multi-bridges}
The scaling theory for multi-bridge polymer networks given in Section \ref{multibridge} can now be applied with the SAW exponents just obtained. As an example, consider the $L$-star boundary configurations as shown in Fig. \ref{fig:DtoNbis}, and imagine that each arm separately obeys a bridge constraint.  We get the configuration exponents 
\begin{eqnarray} 
\gamma_{L,b}(\cdot)=\nu \left[2L-Lx_1^{S}({\rm or})-x_L^{S}(\cdot)\right],
\end{eqnarray}
where $(\cdot)$ represents either $({\rm or})$, or $({\rm o \bullet s})$, or $({\rm sp})$. We thus arrive at the $L$-bridge exponents,
\begin{eqnarray} 
\gamma_{L,b}({\rm or})&=& \frac{9}{32}L\left(3-L\right),\,\,\,
\gamma_{L,b}({\rm o\bullet s})=\frac{3}{32}L\left({13}-3L\right)\\
%\gamma_{L,b}({\rm o\vert s})=\frac{3}{32}L\left({13}-3L\right)\\
\gamma_{L,b}({\rm sp})&=&\frac{1}{32} (51L-9L^2-8).
\end{eqnarray}
For $L=1$, we recover the ordinary single bridge configuration exponent $\gamma_b=9/16$ of Refs \cite{DGK11,DG2019}, and get the new bridge exponents, 
\BE\label{bridgesp}
\gamma_{b}({\rm o\bullet s})=15/16;\,\,\,\, \gamma_{b}({\rm sp})=17/16,
 \EE
at the mixed and special transition points, respectively. 
\subsection{Brownian paths exponents}
Let us simply remark here that for $\kappa=8/3$, the KPZ relations \eqref{eq:KPZ} also encompass the case of two-dimensional Brownian paths, and help calculating  exponents describing various constraints acting on those paths, such as in the case of mutually-avoiding random walks \cite{LW99, D98,BD1999,LSW01a,LSW01b,LSW02,LSWMand,D04,BDLH05}.  For example, the quantum boundary scaling exponent of a single Brownian path avoiding the boundary  is $\Delta^{\mathrm M}_1=1$, and for $L$ mutually-avoiding paths, quantum boundary additivity gives 
\BE\label{deltaLB}
\Delta^{\mathrm M}_L=L\Delta^{\mathrm M}_1=L,
\EE which replaces value \eqref{eq:deltaL} for $L$ mutually-avoiding SAWs. The KPZ relation \eqref{eq:KPZ} then gives  
\begin{eqnarray}\nonumber %\label{eq:KPZLjB}
x^{\mathrm{M},S}_{L}:=\mathcal U_{\kappa=8/3}(\Delta^\mathrm{M}_{L})&=&\frac{1}{3} L \left(2L+1\right), \\ 
%\frac{1}{12}\left(4L^2-1\right),\frac{2}{3} \Delta^\mathrm{M}_{L,j} \Delta^\mathrm{M}_{L,j-1},\,\,\,\,0\leq j\leq L+1,\\ \nonumber
x^{\mathrm{M}}_{L}:=2\mathcal V_{\kappa=8/3}(\Delta^\mathrm{M}_{L})&=&\frac{1}{12}\left(4L^2-1\right),
%2\mathcal U_{\kappa=8/3}\left[\frac{1}{2}\left(\Delta^{\mathrm M}_{L,j+1}\right)\right]\\  \label{eq:BtoBB}
%&=&\frac{1}{3}\Delta^\mathrm{M}_{L,j+1}\Delta^\mathrm{M}_{L,j-1},\,\,\,\,0\leq j\leq L.
\end{eqnarray} 
i.e., the planar Brownian intersection exponents \eqref{eq:MAW}.   

\section{Polymer $\Theta$-point in two dimensions}
\subsection{Introduction}
Let us finally consider polymers at the tricritical $\Theta$-point \cite{Flory,PGG75,PGG79,BauSla2019} in two-dimensions. In Ref. \cite{DS87}, a polymer chain model, inspired by \cite{CJMS87}, of annealed percolating vacancies on the hexagonal lattice, where the percolation transition corresponds to the tricritical $\Theta$-point in two-dimensions, was studied by Coulomb gas methods. A discussion followed about a possible distinction between this so-called $\Theta'$-point model, which is effectively associated with nearest-neighbour and a subset of next-nearest-neighbour contacts, 
 and the usual $\Theta$-point model \cite{Poole88,DS88a,Seno88,DS88b,DS89a,Meir89,DS89b,Seno90,Vanderzande91,Foster92,Stella93}. This was reinforced by the fact that the boundary exponents, $\gamma^{\Theta}_1=8/7$ and $\gamma^{\Theta}_{11}=4/7$ initially proposed in \cite{DS88a}, actually correspond to the \emph{special} transition $\Theta$-point, not the ordinary one, as was finally recognized by Seno, Stella and Vanderzande \cite{Vanderzande91,Stella93}, who further argued that the \emph{ordinary} transition exponents are $\gamma^{\Theta}_1(\mathrm{ord})=4/7$ and $\gamma^{\Theta}_{11}(\mathrm{ord})=-4/7$. A different model was also proposed \cite{Waarnar1992}, but it appears to belong to a different, less stable, universality class \cite{Vernier2015}, and the fact that the $\Theta'$- and $\Theta$-universality classes are the same is nowadays numerically well-verified (see, e.g., \cite{Caracio2011,BGJ19})  and generally accepted. 
 
 \subsection{$Theta$-point multiple chain exponents} Let us now use the results of Sections \ref{DNLQG} and \ref{sec:confweld} to obtain the set of critical exponents of SAWs at their $\Theta$-point in two-dimensions. In the model of Ref. \cite{DS87}, the polymer chains have the same geometrical fractal properties as the hulls of percolating vacancies on the hexagonal lattice. Coulomb gas methods then yield for the percolation hull a fractal dimension $D_{H}=7/4$, hence a correlation length exponent $\nu^\Theta=4/7$ for the polymer chains \cite{HD87,DS87}. Thanks to work by S. Smirnov \cite{Smir01}, the scaling limit of critical percolation on the hexagonal lattice is now rigorously proven to be described by SLE$_{\kappa=6}$.   
% \begin{eqnarray}\label{eq:xSLj}
%x^{S,\kappa}_{L,j}&=& \frac{1}{4\kappa}\left[{2L}+j\,\left({\kappa}-{4}\right)\right]\left[{2L}+(j-1)\,\left({\kappa}-{4}\right)\right],\,\,\,\, 0\leq j\leq L+1,\\ \label{eq:xLj}
%x^{\kappa}_{L,j}
%&=&
%\frac{1}{8\kappa}\left[{2L}+(j+1)\,\left(\kappa-{4}\right)\right] \left[{2L}+(j-1)\,\left(\kappa-{4}\right)\right], \,\,\,\, 0\leq j\leq L.
%\end{eqnarray}
We therefore get from \eqref{eq:xSLj} and \eqref{eq:xLj}
\begin{eqnarray}\label{eq:xSLjt}
x^{S,\kappa=6}_{L,j}&=& \frac{1}{6}\left({L}+j\right)\left({L}+j-1\right),\,\,\,\, 0\leq j\leq L+1,\\ \label{eq:xLjt}
x^{\kappa=6}_{L,j}
&=&
\frac{1}{12}\left({L}+j+1\right) \left({L}+j-1\right), \,\,\,\, 0\leq j\leq L.
\end{eqnarray}
 Specifying to the cases described in Fig. \ref{fig:multiplethetabis} gives
 \begin{eqnarray}\label{eq:KPZLt}
x_{L,j=0}^{\kappa=6}&=&\frac{1}{12}\left(L^2-1\right),\\ \label{eq:BtoBLt}
 x^{S,\kappa=6}_{L,j=0}&=&\frac{1}{6} L \left(L-1\right),\\ 
\label{eq:xSL1t}
  x^{S,\kappa=6}_{L,j=1}&=& \frac{1}{6}L\left({L}+1\right),\\
  \label{eq:xSL2t}
x^{S,\kappa=6}_{L,j=2}&=& \frac{1}{6}\left({L}+1\right)\left(L+2\right).
\end{eqnarray}
Note that in \eqref{eq:KPZLt} and \eqref{eq:xSL1t} we recover precisely the percolation \emph{path-crossing exponents}, first derived in Ref. \cite{AizDupAha99} (see also \cite{BD1999b}).
When considering the largest possible values $j=L+1$ in \eqref{eq:xSLjt} and $j=L$ \eqref{eq:xLjt}, one gets
 \begin{eqnarray}\label{eq:KPZLmax}
x_{L,j=L}^{\kappa=6}&=&\frac{1}{12}\left(4L^2-1\right),\\ \label{eq:BtoBLmax}
 x^{S,\kappa=6}_{L,j=L+1}&=&\frac{1}{3} L \left(2L+1\right),
\end{eqnarray}
which correspond exactly to the planar \emph{Brownian intersection exponents} \eqref{eq:MAW} $x_L^{\mathrm{M}}$ and $x_L^{\mathrm{M},S}$. The reason is that each SLE$_6$ path has for each outer boundary a version of SLE$_{8/3}$, in the same way as a Brownian path hull does, so that they are conformally equivalent here \cite{Mand82,LW00,BD1999,LSW01a,LSW01b,LSW02,LSWMand,D04,BDLH05}. 
%\textcolor{blue}{If we consider, as in Fig.  $2L} SAWs, with 
%\BE\label{eq:xLkrr}
%x_L^{S,\kappa}(\rho_1,\rho_2):=\mathcal U_{\kappa}(\Delta_L^\kappa(\rho_1,\rho_2))=\frac{1}{4\kappa}(2L+\rho_1+\rho_2)(2L+\rho_1+\rho_2+4-\kappa),
%\EE }
  
Recalling the discussion of Sections \ref{conditioning} and \ref{sec:nonsimplepaths} about  the boundary intersection of non-simple SLE$_{\kappa>4}$ curves (Fig. \ref{fig:multiplethetabis}), and applying it to $\kappa=6$, i.e., to $\Theta$-polymer chains, we are led to identify the $j=0$  case as corresponding to the \emph{special} transition, the $j=1$ case to the \emph{mixed ordinary-special} transition, and finally the $j=2$ case to the \emph{ordinary} surface transition of  $\Theta$-polymer chains. This can be summarised as, 
% \begin{eqnarray}\label{eq:KPZLtt}
%&&x^{\Theta}_L=x_{L,j=0}^{\kappa=6}=\frac{1}{12}\left(L^2-1\right),\\ \label{eq:BtoBLtt}
%&&x^{\Theta,S}_L(\mathrm{sp})= x^{S,\kappa=6}_{L,j=0}=\frac{1}{6} L \left(L-1\right),\\ 
%\label{eq:xSL1tt}
%&&x^{\Theta,S}_L(\mathrm{ord\bullet sp})= x^{S,\kappa=6}_{L,j=1}=\frac{1}{6}L\left({L}+1\right),\\
%  \label{eq:xSL2tt}
%&&x^{\Theta,S}_L(\mathrm{ord})=x^{S,\kappa=6}_{L,j=2}=\frac{1}{6}\left({L}+1\right)\left(L+2\right).
%\end{eqnarray}   
 \begin{eqnarray}\label{eq:xt}
&&x^{\Theta}_L=\frac{1}{12}\left(L^2-1\right),\\ \label{eq:xs0t}
&&x^{\Theta,S}_L(\mathrm{sp})=\frac{1}{6} L \left(L-1\right),\\ 
\label{eq:xs1t}
&&x^{\Theta,S}_L(\mathrm{o\bullet s})=\frac{1}{6}L\left({L}+1\right),\\
  \label{eq:xs2t}
&&x^{\Theta,S}_L(\mathrm{or})=\frac{1}{6}\left({L}+1\right)\left(L+2\right).
\end{eqnarray}   
One observes the \emph{shift relations}, $$x^{\Theta,S}_L(\mathrm{o\bullet s})=x^{\Theta,S}_{L+1}(\mathrm{sp}),\,\,\,x^{\Theta,S}_L(\mathrm{or})=x^{\Theta,S}_{L+2}(\mathrm{sp}).$$  
The second shift, $L\mapsto L+2$,  to reach the ordinary case from the special one, is in full agreement with the result of Ref. \cite{Stella93}, $x^{\Theta,S}_{L=1}(\mathrm{ord})=x^{\Theta,S}_{L=3}(\mathrm{sp})=1$. As shown in that work \cite{Stella93}, it is due to the fact that for the $\Theta$-point model of percolating vacancies \cite{DS87}, the vacancies exert an \emph{effective attraction} to the boundary, setting the polymer chains right at the special transition. In the case of an $L$-chain `watermelon', terminally attached to the surface, the ordinary transition can then be mimicked by including its chains inside a single percolating cluster, thus adding the latter's two hull boundaries as two extra $\Theta$-point chains. Finally, note that for $L=2$, the surface duality equation \eqref{eq:dualitysurf} gives for the ordinary surface $\Theta$-point exponent, $x^{\Theta,S}_{L=2}(\mathrm{or})=x^{S,\kappa'=6}_{L=2,j=2}= x^{S,\kappa=8/3}_{L=2,j=0}=2=d$, in full agreement with Eq. \eqref{x2S} and the discussion in Section \ref{Consistency}. 
\subsection{Polymer networks and multi-bridges at the $\Theta$-point} 
The scaling theory developed in Section \ref{multibridge} applies equally well to the $\Theta$-point conditions, and Eqs. \eqref{eq:keyter}, \eqref{eq:keysp}  and \eqref{eq:keytersp}  remain valid, with  $d=2$, and the identifications $\nu\equiv\nu^{\Theta}$, and $x_L\equiv x^{\Theta}_L$ \eqref{eq:xt}), $x_L^S\equiv x^{\Theta,S}_L({\rm or})$ \eqref{eq:xs2t}, and $x_L^S({\rm sp})\equiv x^{\Theta,S}_L({\rm sp})$ \eqref{eq:xs0t}.
\subsubsection{Single chain entropic exponents.}
There is a scaling relation, due to Barber \cite{B73},
\begin{equation}\label{barb}
2\gamma_1 - \gamma_{11} = \gamma + \nu,
\end{equation}
 which  directly follows from \eqref{eq:key}, holds independent of dimension, and is also valid at the $\Theta$-point, both at the ordinary and special transitions.
For terminally attached walks (TAWs) and arches, as well as for single bridges, $L=1,$ and from (\eqref{eq:xt}), \eqref{eq:xs0t} and (\ref{eq:xs2t})  we find $x^{\Theta}_1=0,$ $x_1^{\Theta,S}(\mathrm{or})=1$, and $x^{\Theta,S}_1({\rm sp})=0.$

This gives $\gamma^{\Theta}=\nu^{\Theta}[2-2x^{\Theta}_1]=8/7,$ $\gamma^{\Theta}_1({\rm or})=\nu^{\Theta}[2-x^{\Theta}_1-x_1^{\Theta,S}({\rm or})]=4/7,$ and 
%\begin{equation}\label{g11}
$\gamma^{\Theta}_{11}({\rm or})=\nu^{\Theta}[1-2x_1^{\Theta,S}({\rm or})]=-4/7.$
%\end{equation}
Similarly, at the special transition, we immediately obtain $\gamma^{\Theta}_1({\rm sp})=\nu^{\Theta}[2-x_1^{\Theta}-x_1^{\Theta,S}({\rm sp})]=8/7,$ and 
%\begin{equation}\label{g11sp}
$\gamma_{11}^{\Theta}({\rm sp})=\nu^{\Theta}[1-2x_1^{\Theta,S}({\rm sp})]=4/7.$
%\end{equation}
 These results imply the Barber scaling relation above, and its extension to the exponents at the $\Theta$-point ordinary and special transitions.
 \subsubsection{Single bridge at the ordinary transition.}
\begin{figure}[h!]
\centering
\includegraphics[angle=0,scale =1.32] {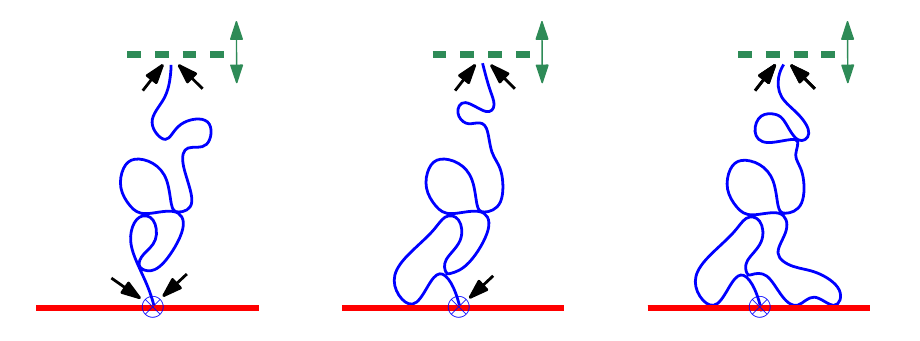}
 \caption{A polymer chain at the $\Theta$-point anchored at a surface, in a bridge configuration in two-dimensions, which corresponds to ordinary boundary conditions as marked by two arrows near the topmost virtual line.  {\sl Left:} Ordinary b.c.'s yield the bridge exponent $\gamma_b^{\Theta}(\mathrm{or})$. {\sl Middle:} Mixed b.c.'s  yield the bridge exponent $\gamma^{\Theta}_b(\mathrm{o\bullet s})$.  {\sl Right:} Special b.c.'s yield the bridge exponent $\gamma^{\Theta}_b(\mathrm{sp})$.}
 \label{fig:singletheta}
\end{figure}
Turning now to bridges, a single bridge at the ordinary transition point can be considered as a TAW rooted at the surface, but with end-point free to move in the bulk, provided it lies in a parallel surface at maximal spacing between the two surfaces (Fig. \ref{fig:singletheta}). We obtain from \eqref{eq:keytersp} that it can be described as a network with ${\mathcal V}=1,$ ${\mathcal V}_S=1,$ $\N=1,$ $n_1=0,$ and $n_1^S=1$, $n_1^b=1$, and all other $L$-vertex numbers vanishing for $L\geq 2$.  This gives the exponent for single ordinary transition bridges as (hereafter dropping $({\rm or})$) 
\begin{equation}\label{eq:gbnug11}\gamma^{\Theta}_b=\nu^{\Theta}[2-2x_1^{\Theta,S}]=\gamma^{\Theta}_{11}+\nu^{\Theta}=0.
\end{equation}
\subsubsection{The special transition.}
For the special transition, the origin vertex has for surface exponent $x_1^{\Theta,S}({\rm sp})$ (Fig. \ref{fig:singletheta}). In this way we find the special bridge exponent,
$$\gamma_b^{\Theta}({\rm sp})=\nu^{\Theta}[2-x_1^{\Theta,S} - x_1^{\Theta,S}({\rm sp})]=\frac{4}{7}.$$ 
From the above, %that we also have $\gamma_{11}({\rm sp})=\nu[1-2x_1^S({\rm sp})]$, and  $\gamma_{11}=\nu[1-2x_1^S]$, so 
we also obtain the special transition scaling relation \cite{DG2019}, 
\begin{equation}\label{eq:gbspg11}
\gamma_b^{\Theta}({\rm sp})=\frac{1}{2}\left[\gamma^{\Theta}_{11}({\rm sp})+\gamma^{\Theta}_{11}\right] + \nu^{\Theta}.
\end{equation}
\subsubsection{Mixed boundary conditions.}
When the attachment vertex is of mixed ordinary-special type, with exponent $x_1^{\Theta,S}({\rm o\bullet s})=1/3$ (Fig. \ref{fig:singletheta}), we get 
\begin{eqnarray}
\gamma_1^{\Theta}({\rm o\bullet s})&=&\nu^{\Theta}[2-x_1^{\Theta} - x_1^{\Theta,S}({\rm o\bullet s})]=\frac{20}{21},\\
\gamma_b^{\Theta}({\rm o\bullet s})&=&\nu^{\Theta}[2-x_1^{\Theta,S} - x_1^{\Theta,S}({\rm o\bullet s})]=\frac{8}{21}.
\end{eqnarray}
\subsubsection{$L$-multi-bridge.} As an application of the
 scaling theory developed in Section \ref{multibridge}, i.e.,  Eqs. \eqref{eq:keyter}, \eqref{eq:keysp}  and \eqref{eq:keytersp},  to $\Theta$-point conditions, consider for exemple the simple case depicted in Fig. \ref{fig:multiplethetabis}.  Assume all chains to be in $\Theta$-point conditions, and all the $L$-bulk single vertices to be local \emph{bridge} vertices. This $L$-bridge configuration gives rise to configurational exponents,
\begin{eqnarray} 
\gamma_{L,b}^{\Theta}(\cdot)=\nu^\Theta \left[2L-Lx_1^{\Theta,S}-x_L^{\Theta,S}(\cdot)\right],
\end{eqnarray}
where $(\cdot)$ represents either $({\rm sp})$, or $({\rm o \bullet sp})$, or $({\rm or})$, depending on the boundary conditions. Eqs. \eqref{eq:xt}, \eqref{eq:xs0t} \eqref{eq:xs1t} and \eqref {eq:xs2t} yield the explicit $L$-bridge exponents,
\begin{eqnarray} 
\gamma_{L,b}^{\Theta}({\rm sp})&=& \frac{2}{21} L(7-L),\,\,\,
\gamma_{L,b}^{\Theta}({\rm o\bullet s})=  \frac{2}{21} L(5-L),\\
\gamma_{L,b}^{\Theta}({\rm or})&=&  \frac{2}{21} (3L-L^2-2).
\end{eqnarray}

\section{Numerical results}
    A number of the scaling relations given above can be compared to recent numerical work in a number of papers. The exponent for bridges at the ordinary transition has been calculated by both series and Monte Carlo methods in two- and three-dimensions by Clisby et al. in \cite{CCG16}. They found $\gamma_b(2\textrm{d})=9/16,$ and $\gamma_b(3\textrm{d})=0.198352 \pm 0.000027.$ These are completely consistent with the scaling relation $\gamma_b = \gamma_{11}+\nu$ given in Eq. (\ref{eq:gbnug11bis}).

    At the special transition, the scaling relation for bridges predicts $\gamma_b(\textrm{sp}) = 17/16$ for the two-dimensional case, and $\gamma_b(\textrm{sp}) \approx 0.746$ for the three-dimensional case. The result for the two dimensional case has been confirmed by a recent series analysis study \cite{BGJ19}. Furthermore, the epsilon expansion, given in \cite{DG2019} gives the values $\gamma_b(\textrm{sp})=1.164$ and $\gamma_b(\textrm{sp})=0.760$ respectively for the exponents in two- and three-dimensions. This is very satisfactory agreement.

For bridges at the theta-point, Beaton et al \cite{BGJ19} found $\gamma_b^{\Theta} = 0.00 \pm 0.03$ which is in perfect agreement with the prediction given in Eq. (\ref{eq:gbnug11}) that this exponent should be precisely zero. For $\gamma_1^{\Theta}$ Seno and Stella \cite{SS88} estimated $\gamma_1^{\Theta} = 0.57 \pm 0.09$ based on a Monte Carlo analysis. Series analysis estimates have been given by Foster et al. in \cite{Foster92} of $0.57 \pm 0.02$ while Beaton et al. \cite{BGJ19} used longer series to estimate  $\gamma_1^{\Theta}= 0.55 \pm 0.03.$ All these estimates are in good agreement with the calculation here that gives $\gamma_1^{\Theta} = 4/7.$ 

For self-avoiding polygons partly adsorbed at the special transition, the situation presents an interesting subtlety.  In two-dimensions, we  find from Eq. \ref{eq:xsurfsp} $x_2^S(\textrm{sp})=1/3,$ and since $\gamma_p(\textrm{sp})=-\nu\, x_2^S(\textrm{sp})$ and $\nu=3/4$, we obtain  the prediction $\gamma_p(\textrm{sp}) = -1/4.$ In \cite{GRW2018} careful series analysis gave the estimate $\widehat \gamma_p \approx -0.754,$ suggesting that the exact value should be $\widehat \gamma_p = -3/4.$

This apparent discrepancy arises due to the way that monomer contacts should be counted in the case of the special transition. If a polygon of $N$ sites has $m$ surface contacts, one can either give the polygon a weight 1 or a weight $m.$ In \cite{GRW2018} a polygon with $m$ monomers in the surface was given a weight 1. In the ordinary-transition case this makes no difference, as when the surface fugacity $a$ is less than the critical fugacity $a_c,$ only a microscopic fraction of the $N$ monomers are in the surface. However at the special transition, one expects a typical value $m\propto N^\phi=N^{1/2},$ because the crossover exponent is exactly $\phi=1/2$ for any  dilute critical $O(n)$ model \cite{BEG89,BE94,FS94}. Thus, introducing the weight $m$, which corresponds to the number of possible choices of the 2-leg root vertex in the partly adsorbed configuration of the polygon,  should change the configuration exponent by $+1/2$. We checked that a reanalysis of the series with weight $m$ yields a configuration exponent  $-0.254$ close to the expected $-1/4$, which accounts for the apparent discrepancy.

Moreover, we remark that Eq. \ref{eq:xLos8/3} gives the \emph{mixed boundary} 2-leg exponent $x_2^S(\textrm{or}\bullet\textrm{sp})=1$, which in turn yields $\gamma_p(\textrm{or} \bullet \textrm{sp})=-\nu\,x_2^S(\textrm{or}\bullet\textrm{sp})=-3/4$.  It is thus tantalising to conclude that the measured value $\widehat \gamma_p \approx -0.754$ in \cite{GRW2018} precisely corresponds to the prediction $\widehat \gamma_p=\gamma_p(\textrm{or} \bullet \textrm{sp})$ instead of $\gamma_p(\textrm{sp})$.  A partly adsorbed polygon configuration, when counted with weight 1, can in fact be associated with that of a polygon with mixed boundary conditions, by viewing the leftmost boundary  vertex of the polygon as a point where the boundary conditions change from {\sl ordinary} to {\sl special}  (or, equivalently, by seeing the rightmost boundary vertex as the point where the boundary conditions change from special to ordinary).

\section{Conclusion}
 We have given several examples showing how the theory of the critical behaviour of $d$-dimensional polymer networks {\cite{BD86,DS86,D89} can be extended to the situation of bridges where the chains lie between two parallel hyper-planes. We have generalized the findings of our earlier work \cite{DG2019} to the case of multi-bridges where several multi-chain vertices are in local bridge configurations.  The exact scaling laws for two-dimensional confined polymer networks made of either self-avoiding walks, or mutually-avoiding walks, or $\Theta$-point walks,  either at the ordinary, or special, or mixed surface transitions, are given. All sets of multiple-path Euclidean scaling exponents are described from a unified SLE perspective, via the canonical use of the fundamental KPZ relation between Euclidean and Liouville quantum gravity exponents.  We also give supportive results based on series and Monte Carlo enumeration data. 
 %Unlike many scaling laws, this requires modification when describing the special transition, and the appropriately modified scaling law is also derived.
 % We have also derived a scaling relation for a subset of SAWs called worms. This has been verified numerically in the two-dimensional case by series analysis. It is also possible to extend the theory  more generally to polymer networks between parallel hyperplanes, {as well as to the case of the tricritical polymer $\Theta$-point,} and this will be the subject of a future article.
\section*{Acknowledgements}
We wish to acknowledge the hospitality of the Erwin Schr\"odinger International Institute for Mathematical Physics where this work was initiated, during the programme on {\em Combinatorics, Geometry and Physics} in June, 2014.  AJG wishes to thank the Australian Research Council for supporting this work through grant DP120100931, and more recently ACEMS, the ARC Centre of Excellence for Mathematical and Statistical Frontiers. We also wish to warmly thank Hans Werner Diehl for pointing out a number of references relevant to surface transitions, and Emmanuel Guitter for his kind help with the figures.

%Otherwise, the following are not cited:
%BBDDG14, C10, C13, C17, Dieh86, DD80, DD81, DieS94 and 98, GLMR, G05, GT84, HM54, HW62, H61, HG94, K13, RG81, R81 so should be deleted.

\end{document}